# 1

# Spectroscopic Needs for Imaging Dark Energy Experiments

Convener: J. Newman

A. Abate, F. Abdalla, S. Allam, S. Allen, R. Ansari, S. Bailey, W. Barkhouse, T. Beers, M. Blanton, M. Brodwin, J. Brownstein, R. Brunner, M. Carrasco-Kind, J. Cervantes-Cota, E. Chisari, M. Colless, J. Comparat, J. Coupon, E. Cheu, C. Cunha, A. de la Macorra, I. Dell'Antonio, B. Frye, E. Gawiser, N. Gehrels, K. Grady, A. Hagen, P. Hall, A. Hearin, H. Hildebrandt, C. Hirata, S. Ho, K. Honscheid, D. Huterer, Z. Ivezic, J.-P. Kneib, J. Kruk, O. Lahav, R. Mandelbaum, J. Marshall, D. Matthews, B. Ménard, R. Miquel, M. Moniez, W. Moos, J. Moustakas, C. Papovich, J. Peacock, C. Park, J. Rhodes, J-S. Ricol, I. Sadeh, A. Slozar, S. Schmidt, D. Stern, T. Tyson, A. von der Linden, R. Wechsler, W. Wood-Vasey, A. Zentner

## EXECUTIVE SUMMARY

Ongoing and near-future imaging-based dark energy experiments are critically dependent upon *photometric redshifts* (a.k.a. photo-$z$'s): i.e., estimates of the redshifts of objects based only on flux information obtained through broad filters. Higher-quality, lower-scatter photo-$z$'s will result in smaller random errors on cosmological parameters; while systematic errors in photometric redshift estimates, if not constrained, may dominate all other uncertainties from these experiments. The desired optimization and calibration is dependent upon spectroscopic measurements for secure redshift information; this is the key application of galaxy spectroscopy for imaging-based dark energy experiments.

Hence, to achieve their full potential, imaging-based experiments will require large sets of objects with spectroscopically-determined redshifts, for two purposes:

- **Training**: Objects with known redshift are needed to map out the relationship between object color and $z$ (or, equivalently, to determine empirically-calibrated templates describing the rest-frame spectra of the full range of galaxies, which may be used to predict the color-$z$ relation). The ultimate goal of training is to minimize each moment of the distribution of differences between photometric redshift estimates and the true redshifts of objects, making the relationship between them as tight as possible. **The larger and more complete our "training set" of spectroscopic redshifts is, the smaller the RMS photo-z errors should be, increasing the constraining power of imaging experiments.**

    ***Requirements:*** Spectroscopic redshift measurements for ~30,000 objects over >~15 widely-separated regions, each at least ~20 arcmin in diameter, and reaching the faintest objects used in a given experiment, will likely be necessary if photometric redshifts are to be trained and calibrated with conventional techniques. Larger, more complete samples (i.e., with longer exposure times) can improve photo-z algorithms and reduce scatter further, enhancing the science return from planned experiments greatly (increasing the Dark Energy Task Force figure of merit by up to ~50%).

*Options:* This spectroscopy will most efficiently be done by covering as much of the optical and near-infrared spectrum as possible at modestly high spectral resolution ($\lambda/\Delta\lambda > \sim 3000$), while maximizing the telescope collecting area, field of view on the sky, and multiplexing of simultaneous spectra. The most efficient instrument for this would likely be either the proposed GMACS/MANIFEST spectrograph for the Giant Magellan Telescope or the OPTIMOS spectrograph for the European Extremely Large Telescope, depending on actual properties when built. The PFS spectrograph at Subaru would be next best and available considerably earlier, c. 2018; the proposed ngCFHT and SSST telescopes would have similar capabilities but start later. Other key options, in order of increasing total time required, are the WFOS spectrograph at TMT, MOONS at the VLT, and DESI at the Mayall 4m telescope (or the similar 4MOST and WEAVE projects); of these, only DESI, MOONS, and PFS are expected to be available before 2020. Table 2-2 of this white paper summarizes the observation time required at each facility for strawman training samples. To attain secure redshift measurements for a high fraction of targeted objects and cover the full redshift span of future experiments, additional near-infrared spectroscopy will also be required; this is best done from space, particularly with *WFIRST-2.4* and *JWST*.

- **Calibration:** The first several moments of redshift distributions (the mean, RMS redshift dispersion, etc.), must be known to high accuracy for cosmological constraints not to be systematics-dominated (equivalently, the moments of the distribution of differences between photometric and true redshifts could be determined instead). The ultimate goal of calibration is to characterize these moments for every subsample used in analyses - i.e., to minimize the *uncertainty* in their mean redshift, RMS dispersion, etc. - rather than to make the moments themselves small. Calibration may be done with the same spectroscopic dataset used for training **if** that dataset is extremely high in redshift completeness (i.e., no populations of galaxies to be used in analyses are systematically missed). **Accurate photo-z calibration is necessary for all imaging experiments.**

    *Requirements*: If extremely low levels of systematic incompleteness ($< \sim 0.1\%$) are attained in training samples, the same datasets described above should be sufficient for calibration. However, existing deep spectroscopic surveys have failed to yield secure redshifts for 30-60% of targets, so that would require very large improvements over past experience. This incompleteness would be a limiting factor for training, but catastrophic for calibration. If $< \sim 0.1\%$ incompleteness is not attainable, the best known option for calibration of photometric redshifts is to utilize cross-correlation statistics in some form. The most direct method for this uses cross-correlations between positions on the sky of bright objects of known spectroscopic redshift with the sample of objects that we wish to calibrate the redshift distribution for, measured as a function of spectroscopic $z$. For such a calibration, redshifts of ~100,000 objects over at least several hundred square degrees, spanning the full redshift range of the samples used for dark energy, would be necessary.

    *Options:* The proposed BAO experiment eBOSS would provide sufficient spectroscopy for basic calibrations, particularly for ongoing and near-future imaging experiments. The planned DESI experiment would provide excellent calibration with redundant cross-checks, but will start after the conclusion of some imaging projects. An extension of DESI to the Southern hemisphere would provide the best possible calibration from cross-correlation methods for DES and LSST.

**We thus anticipate that our two primary needs for spectroscopy - training and calibration of photometric redshifts - will require two separate solutions.** For ongoing and future projects to reach their full potential, new spectroscopic samples of faint objects will be needed for training; those new samples may be suitable for calibration, but the latter possibility is uncertain. In contrast, wide-area samples of bright objects are poorly suited for training, but can provide high-precision calibrations via cross-correlation techniques. Additional training/calibration redshifts and/or host galaxy spectroscopy would enhance the use of supernovae and galaxy clusters for cosmology. We also summarize additional work on photometric redshift techniques that will be needed to prepare for data from ongoing and future dark energy experiments.

## 1.1 Introduction: The problem of photometric redshift calibration

The principal effect of dark energy is to alter the expansion history of the universe, particularly by acceleration at late times. It also affects the rate at which cosmic structures grow. Therefore, in order to obtain tight constraints on the effects of dark energy, it is crucial to have measurements of the redshifts of galaxies. The redshifts effectively provide a third, radial dimension to studies of cosmology, enabling the use of three-dimensional information and to study phenomena as a function of distance and time (which are both monotonically related to the redshift, $z$). This additional information also enables the identification of large structures (such as galaxy clusters) in three-dimensional space.

Redshifts may be measured with precision via spectroscopy. However, it is completely infeasible with either current or near-future instruments to obtain redshifts in this matter for the vast numbers of faint galaxies (of order a few billion) that will be utilized in the next generation of dark energy experiments. Therefore, upcoming imaging-based projects such as the Dark Energy Survey (DES), HyperSuprimeCam (HSC), *Euclid*, *WFIRST-2.4*, and the Large Synoptic Survey Telescope (LSST) need an alternative avenue to provide a fully three-dimensional picture of the distribution of galaxies and to trace the expansion of the Universe and growth of structure as a function of time.

As a result, these experiments will primarily rely on *photometric redshifts* -- i.e., estimates of the redshift (or the probability distribution of possible redshifts, $p(z)$) for an object based only on imaging information, rather than spectroscopy (Spillar 1985, Koo 1999). Effectively, multi-filter imaging provides a very low-resolution spectrum of an object, which can be used to constrain its redshift. Because flux from a relatively wide wavelength range is being combined for each filter, imaging provides a higher signal-to-noise ratio than spectroscopy; however, broader filters provide cruder information about an object's spectrum, and hence on $z$ or $p(z)$.

A wide variety of techniques for estimating photometric redshifts have been developed, ranging from the most straightforward -- determining the redshift that minimizes the $\chi^2$ or maximizes the likelihood when some set of empirical or synthetic template galaxy spectra are fit to the observed multiband photometry of a galaxy -- to Bayesian methods, to neural-network or other machine-learning based estimators that rely on having a training set that matches the galaxies to which the method will be applied (e.g., Budavari 1999; Collister & Lahav 2004). In current samples, a wide variety of techniques offer very similar performance (Hildebrandt et al. 2010).

Photometric redshift algorithms can also provide additional information about target galaxies that is useful for identifying the best samples for dark energy studies. For instance, intrinsically red 'early-type' galaxies, which may be identified via their colors, possess less dust than other types. As a result, extinction – currently one of the dominant sources of systematic uncertainty - should be less of an issue for Type Ia supernova studies if samples are restricted to only SNe in early-type hosts. We refer the reader to Lahav, Abdalla and Banerji (2008) for a more detailed review of photometric redshift techniques and algorithms.

In this white paper, we focus on estimating the spectroscopic requirements for tuning and calibrating photometric redshift algorithms for ongoing/near-future ("Stage III") and next-generation ("Stage IV") dark energy experiments, as well as laying out the open questions that remain for determining those requirements with precision.

## 1.2 Twin needs for spectroscopy: Training and Calibration

We can divide the needs of spectroscopic redshifts for imaging dark energy experiments into two broad classes, which we will refer to in this white paper as "training" versus "calibration;" in general, calibration will be the harder problem.  ***Training*** constitutes the use of samples with known redshift to develop or refine algorithms, and hence to reduce the error on each photometric redshift estimate. In general, machine-learning based techniques will rely entirely on having a sample of objects with known redshift which is used to map the relationship between galaxy color and $z$; this limits the ability of these methods to extrapolate beyond the bounds of this training sample. Methods that instead rely on physical models of galaxy spectra still rely on training sets of galaxies to develop and refine spectral templates and flux calibrations and to determine priors on redshift distributions. The better / more complete the training sample of spectroscopic redshifts is, the more accurate the photometric redshift estimates for individual objects will be, which improves constraints on dark energy (see Figures 1-1 and 1-2). Essentially, **the goal of training is to minimize all moments of the distribution of differences between photometric redshift and true redshift**, rendering the correspondence as tight as possible, and hence maximizing the three-dimensional information in a sample.

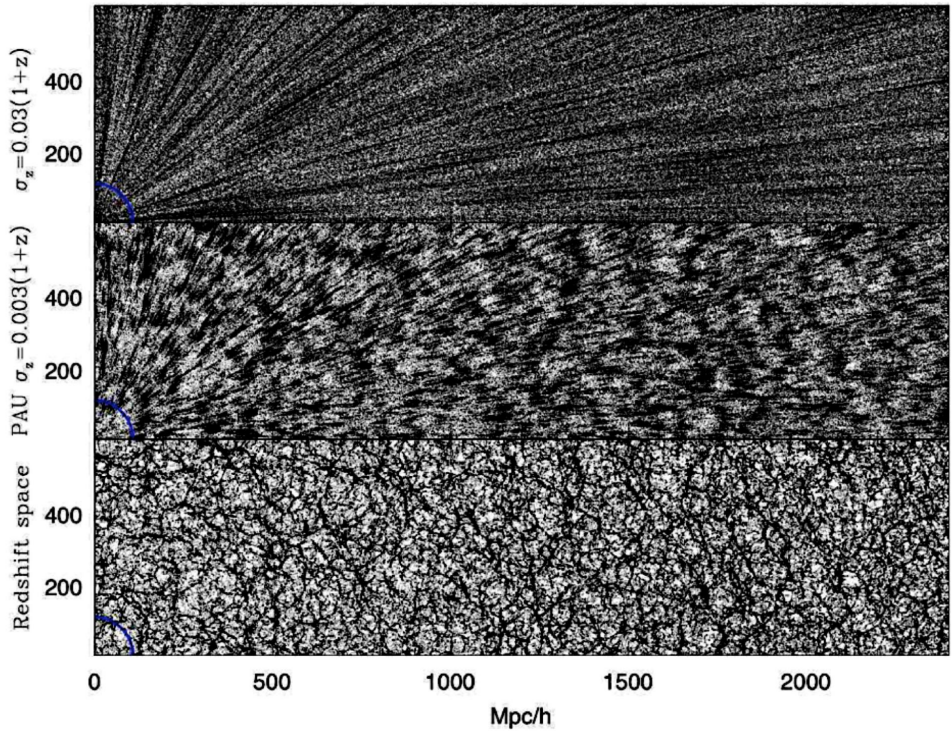

**Figure 1-1.** *A demonstration of the benefits of improved photometric redshift accuracy via better spectroscopic training sets, adopted from Benitez et al (2009). The bottom panel shows a simulated slice of the galaxy density field, with galaxies' true (spectroscopic) redshifts used to set the distance from the origin. The middle and top panel show the spatial distribution of galaxies as observed in the presence of increasingly larger photo-z errors, with RMS of $\sigma_z=0.003(1+z)$ or $0.03(1+z)$, respectively, where z is the true redshift of a given galaxy. Without improved training sets, LSST photometric redshift errors should have RMS $\sim 0.05(1+z)$, worse than the top panel. However, based on tests with simulations, LSST imaging may be capable of delivering errors of $\sim 0.005(1+z)$ for objects at z<1.4 if sufficiently high-quality training sets are available, delivering performance comparable to the middle panel of the figure for that critical redshift range.*

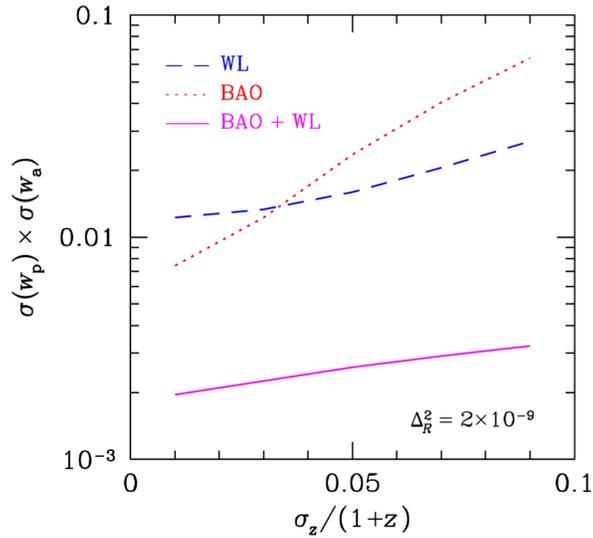

**Figure 1-2.** *Impact of improving photometric redshift errors on dark energy inference (from Zhan 2006). The plot shows the product of errors on two parameters describing the dark energy equation of state ($w = \frac{P}{\rho}$) - its value at the best-constrained redshift, $w_p$, and its derivative with respect to scale factor of the universe, $w_a = \frac{dw}{da}$ - as a function of the RMS error in the photometric redshifts used in the analysis, for a variety of cosmological probes. Constraints on dark energy from weak lensing (WL) improve only slowly as photo-z errors improve, while the BAO constraints, which rely on characterizing features in the large-scale structure seen in Figure 1-1, benefit greatly; the combination has intermediate behavior. Without additional training sets, LSST performance should be $\sigma_z=0.05(1+z)$, which is sufficient to deliver dark energy measurements well within the range expected for a Stage IV experiment; however, improved photo-z errors would increase the constraining power further. Based on tests with simulations, the LSST system is capable of delivering errors of $\sigma_z=0.02(1+z)$ to z=3 if perfect template information is available; simulated performance is even better for z<1.4, achieving $\sigma_z\sim0.005(1+z)$ Achieving that would improve the Dark Energy Task Force figure of merit from the experiment by ~50%.*

The problem of **calibration** is that of determining the overall redshift distribution of samples of objects that are selected based upon their properties (most often, that property will be their estimated photometric redshift). In general, while individual photometric redshifts need not be known with precision for most dark energy science, the *aggregate* properties of samples selected according to photo-z information must be determined with very high accuracy; hence, high-precision, robust calibration is necessary. If photo-z's are systematically biased in an unknown way or if their errors are poorly understood, dark energy inference will be biased as well (e.g., because we are effectively measuring the distance to a different redshift than is assumed in calculations). Essentially, **the goal of calibration is to determine with high accuracy the moments of the distribution of true redshifts of dark energy samples.**

Most dark energy analyses will rely on dividing objects up into subsets or bins in some way (typically, according to photometric redshift and/or galaxy properties such as color or size); the redshift distribution for each bin must then be well-characterized, typically quantified by the first two moments. For instance, for LSST, it is estimated that the mean redshift, $<z>$, and the $z$ dispersion, $\sigma_z$, for samples of objects in a single photo-z bin must be known to $\sim 2\times 10^{-3}$ $(1+z)$ or better for dark energy inference not to be systematically degraded (Zhan & Knox 2006, Zhan 2006).

The impact of uncertainties in $<z>$ and $\sigma_z$ on dark energy constraints is illustrated in Figure 1-3. Requirements for Stage III surveys will be somewhat less stringent. Past a certain point, larger samples of spectroscopic redshifts are more beneficial for calibration than training: uncertainties in photometric measurements and degeneracies between possible redshifts will provide fundamental limits to the precision with which photometric redshifts may be determined, but uncertainties in $<z>$ and $\sigma_z$ will still decrease as the square root of the number of spectra (or number of independent regions of sky sampled, if field-to-field variations dominate). Accurate calibration is a necessity for any photometric redshift estimates used for dark energy studies, regardless of whether they are template- or training-set based.

For both training and calibration purposes, we require a sample of objects for which the true redshift is known; this information can only be obtained securely via spectroscopic measurements. In Section 2 of this white paper, we will discuss the design parameters for spectroscopic training sets that will allow Stage III and IV imaging dark energy experiments, including DES, HSC, and LSST, to reach their full potential.

If secure spectroscopic redshifts could be obtained for a sufficiently large fair sample of those objects to which photometric redshifts will be applied, both training and calibration needs can be fulfilled by the same set of objects. However, real spectroscopic samples have fallen well short of this goal (see §2); hence photo-$z$ calibration is likely to be determined using cross-correlation-based techniques, which combine information from multiple datasets, taking advantage of the fact that many observables each trace the same underlying dark matter distribution. Those methods are discussed extensively in a separate Cosmic Frontiers white paper focused on many aspects of cross-correlations; in Section 3 of this paper, we summarize the spectroscopic requirements for such work. In Section 4 of this white paper, we present a variety of avenues for future work that can increase the power of photometric redshifts in future dark energy experiments. Finally, in section 5 we summarize our conclusions.

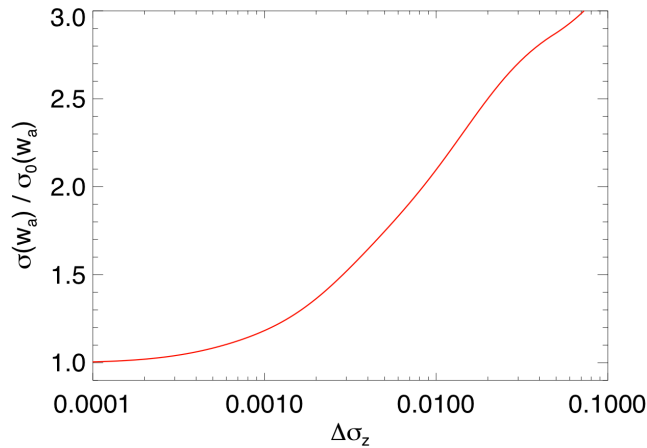

**Figure 1-3**. *Impact of uncertainty on the mean redshift ($\Delta<z>$) or the RMS of the redshift distribution ($\Delta\sigma_z$) on dark energy constraints. The y-axis shows the relative increase in the error in $w_a = \frac{dw}{da}$, the derivative of the equation of state (pressure-to-energy ratio) of dark energy with respect to scale factor a, resulting from an LSST-like weak lensing analysis, as a function of the uncertainty in the RMS of each $\Delta z = 0.1$ wide bin used in the analysis. For Gaussian statistics, the error in the mean redshift, $\Delta<z>$, will be larger than the plotted x coordinate value by a factor of $\sqrt{2}$; the impact of both of these errors in concert are included in this analysis (based on the work of Hearin et al. 2011). Uncertainties of a few hundredths in z in the mean redshift and RMS of galaxy samples, as has typically been achieved in the past, will not be sufficient for future surveys, as it leads to a factor of a few degradation in errors on this crucial dark energy parameter.*

## 2.1 Training sets for photometric redshift algorithms

In this subsection, we consider a variety of factors which determine the characteristics of spectroscopic samples that will be used for photometric redshift training should have, and summarize the greatest challenges we may expect to encounter. Because photometric redshift precision will ultimately be limited by the variation in and degeneracies between spectral energy distributions amongst all galaxies and how that range depends on redshift, factors which are all currently unknown, it is difficult to predict at what point the training information in spectroscopic datasets will saturate. However, it is more straightforward to determine the scope of spectroscopic datasets required for conventional, direct calibration of photometric redshift errors and redshift distributions, as we might hope to be possible with our training spectroscopy. Since it is anticipated that calibration requirements will be somewhat more difficult to meet than training needs, calibration needs then determine the minimum size of training sets required. In the following subsection (§2.2), we will estimate the resources needed to obtain the requisite spectroscopic training data, guided by the considerations presented here.

***Required training sample sizes***   If the distribution of the difference between photometric redshifts and true (spectroscopic) redshifts is assumed to be a Gaussian, the standard deviation and mean of this distribution as a function of redshift need to be known to better than ∼0.002(1+$z$) precision in order to not degrade dark energy parameter accuracies from Stage IV surveys by more than 50% (cf. Ma et al. 2006, Huterer et al. 2006, Kitching et al. 2008, Bernstein & Huterer 2010, Hearin et al. 2010, Hearin et al. 2012). Naively, this necessary calibration accuracy would be simple to achieve. The error in the mean of a sample with standard deviation $\sigma$ as determined from N samples will be $\sigma/\sqrt{N}$, while the uncertainty in the standard deviation is $\sigma/\sqrt{2N}$, so for photometric-redshift errors of ∼0.05(1+$z$), only a few hundred calibration redshifts would be required per redshift bin to achieve sufficient accuracy. With 10-20 redshift and/or parameter bins total, less than 5000 calibration (and hence training) redshifts would be required, presuming that spectroscopic redshifts could be obtained for a fair sample out of all objects within each bin (we ignore the impact of sample/cosmic variance here; we discuss that below instead).

If the photometric redshift error distribution cannot be described by a single Gaussian, the requirements placed on calibration samples become more stringent (Ma & Bernstein 2008). For example, requiring characterization of catastrophic outliers can cause the estimated required spec-z sample size to become as large as $10^6$ in order for dark energy parameters not to be strongly degraded. However, if the analysis is restricted to only include objects estimated to lie at $z$<2.5, the size of the spec-z sample could be reduced to just ∼$10^4$, a result of excluding low-z galaxies erroneously identified as high-z ones (Bernstein & Huterer 2010); this would modestly degrade the constraining power of the sample (by removing information at high $z$) in exchange for a more robust calibration.

In Figure 2-1 we illustrate how the dark energy constraints improve as calibration samples are enlarged. This figure is analogous to Figure 1-3, but here we quantify the prior knowledge on $\sigma_z$ and $z_{bias}$ in terms of the number of galaxies in the calibration sample $N_{spec}$. However, in this case we assume that prior information is apportioned between different bins in redshift according to the overall redshift distribution $n(z)$, rather than via the simple $\sigma/\sqrt{N}$ scaling relation described above; thus at fixed $N_{spec}$, the constraints shown in Figure 2-1 are in general weaker than they would be if prior information were constant in redshift. To facilitate a direct comparison between DES and LSST, we have plotted the relative degradation in the statistical constraints on $w_{piv}$, that is, the value of the dark energy constraint at the redshift where it is best constrained by the survey.

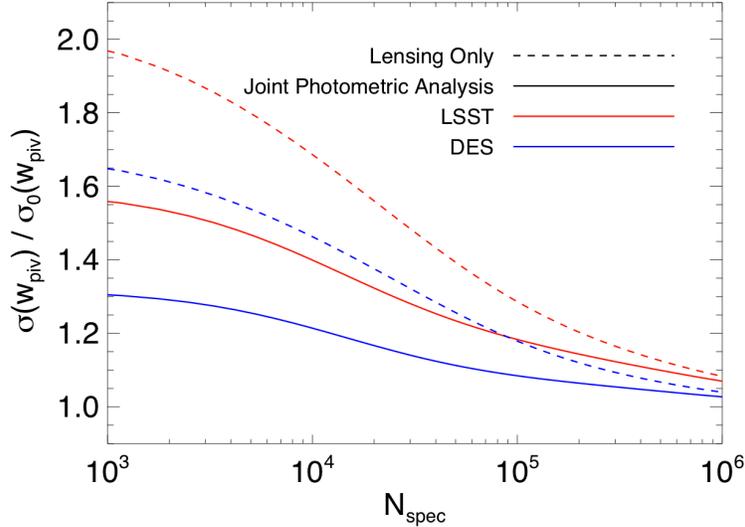

**Figure 2-1.** *Scaling of the errors on $w_{piv}$, the dark energy equation of state parameter at its best constrained redshift, with $N_{spec}$, the number of galaxies in the spectroscopic calibration sample, based on the work of Hearin et al. 2011. Each curve is normalized by the error from statistical uncertainties in the limit of perfect knowledge of the true probability distribution function of redshift as a function of estimated photo-z. Dashed curves pertain to constraints attainable with a weak lensing-only analysis, and solid curves to the case where large-scale galaxy clustering in the imaging data has been utilized as well. The LSST constraints degrade more rapidly with weaker priors than those for DES, reflecting the more stringent calibration demands of this survey. For LSST, the constraining power of cosmic shear is weakened by 35% for a calibration sample size of $3 \times 10^4$ compared to a perfectly-calibrated dataset; this degradation is less than 15% when information from galaxy clustering is incorporated.*

The dashed curves in Figure 2-1 illustrate the ability of each of these surveys to constrain dark energy with a cosmic shear (weak lensing)-only analysis. As is clear from this figure, LSST places stronger demands on the photo-z calibration program than DES, due to its smaller intrinsic statistical errors. With 30,000 total spectra used in the calibration, the dark energy constraints are degraded by ∼35% relative to the limit of perfect prior knowledge of the photo-z distribution if only information from lensing is used.

Figure 2-1 also demonstrates one benefit of including cross-correlations between gravitational lensing shear and galaxy clustering. The solid curves labeled "joint photometric analysis" pertain to analyses of imaging data that employ cosmic shear statistics jointly with galaxy clustering measurements in the linear regime (∼degree scale correlations). The improvement that comes from including clustering information is considerable, especially in the regime of low $N_{spec}$. The application of cross-correlation information to the photo-z calibration program is explored more fully in a companion white paper.

For several reasons, the above estimates on the number of required calibration spectra (and hence on the minimum size of our training sample) are only approximate. For example, the broader the distribution of colors used in the lensing analysis, the larger the number of spectra that are required in order to faithfully sample from the galaxy distribution. Since the number density of galaxies in a particular region of color space is significantly uncertain at high redshift ($z > \sim 1$), exact estimates of the calibration requirements are not possible with our current knowledge of the high redshift universe.

Additionally, uncertainty in our ability to predict the matter power spectrum in the nonlinear regime translates into concomitant uncertainty in the photo-z calibration requirements (e.g., Hearin et al. 2010,

Hearin et al. 2012). This degeneracy has a simple geometric origin: at fixed angular scale, the image distortions of more distant galaxies receive a greater contribution from larger scale matter over-densities relative to galaxies at lower redshift. Thus, the more precisely a galaxy's redshift is known, the narrower the range of scales in the cosmic density field that contribute to the lensing signal. Therefore, the required photo-$z$ precision depends on our uncertainty in the amount of power in the density field. Since our ability to accurately predict the matter power spectrum on scales relevant to lensing is still limited, this effect is not small: the requirements on our knowledge of the photo-$z$ dispersion can vary by a factor of ~5 between commonly-used fitting formulae for the matter power spectrum (though some of this variation may be due to known issues in HALOFIT).

*Scope of current and near-future datasets* Existing spectroscopic samples will be useful for photometric redshift training and calibration, but fall far short of the needs for even Stage III experiments. The $i$=23.7 faint magnitude limit of the Dark Energy Survey matches the depth limit of the VVDS DEEP sample; however, that dataset contains only a few thousand secure (>95% confidence, i.e., flag 3 or 4) redshifts in the magnitude range $22.5 < i < 24$, split over only two fields (Le Fevre et al. 2005); this falls far short of DES calibration needs. VVDS WIDE has roughly 9,000 secure redshifts to $i$=22.5, roughly 4 times brighter flux than the DES magnitude limit, almost all in a single region of the sky. DES will also overlap with two fields containing more than 17,000 secure redshifts from the DEEP2 Galaxy Redshift Survey (Newman et al. 2013), but in those fields DEEP2 rejected $z < 0.7$ objects from the spectroscopic sample, limiting the redshift coverage. The $R$=24.1 magnitude limit of DEEP2 is equivalent to $i$=24 for the bluest galaxies, but only $i$~22.5 for the reddest.

Although not in the DES footprint, the COSMOS field covered by the zCOSMOS survey (Lilly et al. 2009) is accessible to the Blanco telescope and could be covered by a separate imaging campaign to facilitate photo-$z$ calibration; zCOSMOS-bright has ~11,000 secure redshifts to $i$=22.5, while zCOSMOS-deep will provide ~6,000 redshifts for fainter objects; however, it will reach the DES limit for blue objects at redshifts $z$>~1.5, but not in general due to the $B$-band limit and color pre-selection applied. The VIPERS survey (Guzzo et al. 2013) will provide tens of thousands of secure redshifts for a color-selected subset of $i$<22.5 objects spread out over 24 square degrees in 2 fields that overlap DES. The PRIMUS survey (Coil et al. 2011) has obtained secure redshifts for roughly 80,000 objects over 7 fields overlapping with DES, totaling 9.1 square degrees; but the sample is highly incomplete beyond $i$=22. A variety of other surveys (such as SDSS (Ahn et al. 2012), BOSS (Dawson et al. 2012), WiggleZ (Parkinson et al. 2012), and GAMA (Driver et al. 2011)) also overlap with DES, but target objects even more bright compared to the survey depth, limiting their usefulness for photometric redshift calibration. Although we have focused specifically on DES, the situation is similar for HyperSuprimeCam surveys, with better overlap with DEEP2 and zCOSMOS but less with VIPERS or WiggleZ.

Putting all this together, we can see that samples of the desired scale in terms of number of objects exist; but those large samples are much brighter than the samples needed for photometric redshift training/algorithm development for Stage III surveys. One small, significantly incomplete sample exists to the DES survey limit. The situation is worse for Stage IV surveys; the LSST lensing "gold sample" extends to $i$=25.3, while the *WFIRST* satellite will require spectroscopic redshifts for samples magnitude-limited at infrared wavelengths (which can correspond to very faint optical magnitudes and hence very poor redshift success from the ground; see, for instance, the results of the Gemini Deep Deep Survey (Abraham et al. 2004).

**It is therefore clear that new, deeper spectroscopic samples will be required if we wish to optimize photometric redshift methods for Stage III and Stage IV projects.** The existing spectroscopic datasets will continue to be useful for training photometric redshifts, but they are not sufficient for the needs. The new spectroscopy required will benefit greatly from the lessons learned from the

previous generation in both experiment and instrument design, as well as being able to take advantage of software and human expertise developed through the previous generation of spectroscopic surveys.

We can anticipate that future projects that have been primarily designed to study galaxy evolution will expand the set of available redshifts, particularly in terms of depth. For instance, 300 nights in total have been allocated by Subaru for surveys with the PFS spectrograph, which should begin around 2018. Current plans for this time include spectroscopy of color-selected samples to a depth of $J_{AB}$ ~ 23.5 (roughly corresponding to the $i$~23.7 depth of DES imaging) over an area of ~10 deg$^2$, as well as a broadly-selected sample of 5000 objects to similar depth, covering ~3 deg$^2$. The resulting set of redshifts will be of significant use for training photo-z algorithms and could potentially contribute to calibration samples, though it will only cover a fraction of the needs for stage III and IV experiments.

***The problem of incompleteness***   Based on experience with the existing spectroscopic datasets, we can anticipate that obtaining *statistically complete* (i.e., fair-sample) training sets to the depths of Stage III or Stage IV surveys will be extremely challenging. In actuality, the surveys described above obtain secure redshifts for only those targeted objects which exhibit multiple strong spectral features within the wavelength window covered by spectroscopy; it is far from a random subset of targets. The fraction of targets that fail to yield secure redshifts is large. For instance, the DEEP2 Galaxy Redshift Survey, for redshift quality classes with >99.5% (vs. >97.5%) reproducibility, obtained ``secure'' redshifts for only 60% (vs. 75%) of the >50,000 galaxies targeted (Newman et al. 2012). Other surveys have done worse, with high-confidence redshift measurements for 21-59% of targeted galaxies (Le Fevre et al. 2005, Garilli et al. 2008, Lilly et al. 2009), depending on the survey and whether >95% or >99% secure redshifts are required. Surveys of fainter galaxies have even higher rates of failure (Abraham et al. 2004). Achieving even this limited level of redshift success in deep spectroscopic samples has been labor-intensive, requiring checks by eye of each spectrum.

The LSST `gold sample' for dark energy measurements extends to $i$=25.3, more than 10 times fainter than those objects for which redshifts may be obtained (with ~60-75% completeness) at 8-10m diameter telescopes in one hour of observing time (Lilly et al. 2009, Newman et al. 2012). Assuming that spectra of the same signal-to-noise should yield the same redshift success rate, this implies that least 100 hours of observation time would be required to achieve the same 60-75% success rate for LSST calibration redshifts to $i$=25.3, or 5 hours of observation time at DES depth ($i$=23.7). Depending on the multiplexing capabilities of the spectrograph used, anywhere from a few dozen to a few thousand spectra would be obtained from each observation. Doing better is extremely expensive in telescope time; based on the results of existing surveys, we anticipate that obtaining >90% redshift completeness for red galaxies at the LSST magnitude limit would take roughly one thousand hours of observation time (more than 100 nights) on an 8–10m telescope. Deep infrared spectroscopy from space could lessen these problems, but will not solve them, due to the small field of view (and hence high shot noise and sample variance) of JWST and the limited depth, spectral resolution, and wavelength range of *WFIRST-2.4* or *Euclid* spectroscopy.

With existing instruments, at most 100-300 redshifts can be obtained at a time, while the proposed PFS spectrograph (Ellis et al. 2012) should be able to target >2000 objects at once; hence obtaining 75% complete spectroscopic samples for tens of thousands of objects would require a large investment of telescope time. Even a telescope with a 30m diameter mirror would require >10 hours to obtain 75%-completeness redshifts to LSST depth, and would do so for only a limited number of objects in a small field of view. **Obtaining spectroscopic training datasets to Stage IV depth will require a major international effort**, especially so if these data will be used for precision calibration of photometric redshifts. The global scope of this work is reflected in the set of co-signers of this document. These efforts will require a long lead time to spread telescope-time allocations over many years, so must begin soon. Fortunately, the same set of spectroscopy can provide benefits to many different imaging surveys, as they will all overlap in the galaxy populations utilized. Additional multi-wavelength imaging has the potential to bootstrap information from training sets to broader samples and mitigate some of the effects of incompleteness (as discussed in §4-1); obtaining that imaging, too, will require investments of telescope time from a wide variety of sources spanning multiple years.

We note that if the sorts of galaxies that fail to yield redshifts are localized in color space, problem galaxies could be excluded from dark energy analyses at the cost of reduced sample size; however, this is unlikely to always be possible, and could weaken dark energy constraints, as described below. For LSST and other Stage IV surveys, we will need to be prepared for the great likelihood that we will have only incomplete training sets, both in considering photo-$z$ algorithms and in establishing calibration plans.

The effects of incompleteness on photometric redshift training are difficult to predict for template-based methods. However, for training-set/machine-learning-based techniques, the impact is strong and direct; objects missing from training sets will result immediately in biased photometric redshift distributions with, typically, smaller estimated scatter than the true value (as only objects at certain redshifts will be included in the training set and used to predict redshift distributions). Some machine-learning-based techniques (e.g., neural networks) are highly nonlinear and tend to extrapolate poorly; their performance with systematically incomplete training sets requires further investigation for Stage III and IV surveys.

Stage III surveys will need to make progress on the utilization of incomplete training and calibration spectroscopy, as it is already a major issue at DES depth. However, Stage IV photo-$z$ calibration requirements are ~two times more stringent, so additional efforts will be necessary. It has been estimated that systematic incompleteness in conventional calibration samples must be <~0.1% if we are to ensure that dark energy constraints from future experiments will not be degraded even in worst-case scenarios (Bernstein & Huterer 2010). If the true situation is favorable - with objects that would fail to yield redshifts being identifiable based on photometric properties, and objects that are placed in the wrong photometric redshift bin due to training set incompleteness being in general counterbalanced by exchanges between bins in the opposite direction - then it is estimated that systematic redshift failure rates as high as 1% may be acceptable (Cunha et al. 2012b); however, it is unknown if these conditions will be met. Regardless, existing deep samples fall far short of even this goal; we must proceed with the expectation that other methods of calibration may ultimately be necessary (see §3.1). Although it is a critical issue for calibration, redshift incompleteness will only degrade, rather than compromise, photometric redshift training. Incomplete training sets will still be highly valuable for improving photo-z accuracy even if they prove to be insufficient for Stage III and IV calibration needs.

***Minimizing systematic errors via restricted samples*** Given the limited number of broad-band filters available from DES or LSST photometry, and even including near infrared bands from *Euclid* or *WFIRST-2.4*, inferences of the redshifts of galaxies will be limited by unavoidable degeneracies in the problem. Even if we had complete and representative training sets and extremely high signal-to-noise data, there are regions of color/flux parameter space that are consistent with multiple redshifts. Existing photo-$z$ algorithms often provide a redshift confidence or redshift probability distribution that can be used to identify the most likely outlier candidates, for instance those galaxies which have multiple potential redshift solutions with significant probability; see, for instance, Benitez 2000, Oyaizu et al. 2008, and Gorecki et al. 2013. Alternatively, regions of parameter space with known problems can be wholly discarded (e.g., Jain et al. 2007, Nishizawa et al. 2010). Either type of cut decreases the galaxy sample size; the severity of the cuts may be adjusted to balance needs of sample size versus photo-$z$ quality depending on the particular science goals. They must be utilized with care, however; for instance, regions with lower-quality data may tend to have less-certain photometric redshifts, causing objects to be excised in a spatially-dependent way that will imprint a false signal in large-scale structure measurements. Such cuts would also be likely to cause systematic incompleteness in the populations of galaxies included as a function of redshift, as degeneracies will occur for galaxies of only certain intrinsic colors; this can complicate certain analyses.

If galaxies are eliminated from the sample by such cuts equally at all redshifts, the dark energy constraints will generally degrade by a constant factor (or not at all, if sample/cosmic variance - q.v. below - dominates errors). However, it will more commonly be the case that there are limited redshift ranges where degenerate solutions are possible. In that case, the impact on dark energy inference from photo-z quality cuts may be weak or strong depending upon the redshift range affected. The greatest potential systematic errors on dark energy will arise from confusing objects at the lowest and highest redshifts in the sample; fortunately, cutting the photometric sample at very high- and low-redshift has only mild consequences for the statistical constraining power even for stage IV surveys such as LSST.

As shown in Figure 2-3 (adapted from Hearin et al. 2010), if an LSST sample is restricted to the range $0.2 < z_{ph} < 2$, the achievable statistical constraints on the current value of the dark energy equation of state parameter, $w_0$, only degrade by ~20%. Such a photo-$z$ cut is coarse, but Fig. 5 shows that it would come at a relatively low cost. Since systematics caused by misidentifying galaxies with very low/high redshifts are severe (and not uncommon, e.g., 4000Å/Lyman break confusion), such cuts will be very useful for ensuring that redshift distributions are robust and for minimizing potential photo-z systematics.

The problem of catastrophic outliers is exacerbated if incomplete training+calibration datasets do not enable us to identify problematic regions of parameter space. For instance, if a class of galaxies is missing in the spectroscopic sample, they may be placed in an incorrect redshift bin based on other objects with similar colors but successful redshifts. If the fraction of such mis-identified galaxies exceeds ~$10^{-3}$, constraints on dark energy can be weakened significantly (e.g., Hearin et al. 2010). Methods of identifying and (if there are no systematically missed populations) correcting regions with incomplete/unreliable training sets have been explored in Lima et al. 2008, Cunha et al. 2009, Cunha et al. 2012b, and Carrasco-Kind & Brunner 2013a. The use of cross-correlation techniques (Newman 2008, Matthews & Newman 2010) will enable us to identify regions of photo-$z$ and/or color space that exhibit catastrophic outliers, which potentially can then be excluded from dark energy studies.

***Optimizing selection of objects for spectroscopy*** In principle, it should be possible to select galaxies for spectroscopy in an optimal manner in order to obtain the minimal set that determines the mapping from properties to redshift with the desired accuracy of both training and calibration. However, it remains unclear how small this subset may be (we have employed conservative estimates above) and how best to select it. Simply choosing a random subsample of all photometric objects with matching parameter distributions (in flux, size, etc.) will only sparsely sample or miss rare objects in the tails of the distribution, for instance; while dark energy constraints are most sensitive to the bulk of objects, which will be found in the core of the photo-z distribution (see, e.g., Cunha et al. 2012a). Carrasco-Kind & Brunner (2013a) have shown that random sampling is inefficient at improving the accuracy of photometric redshift estimation, but that targeted spectroscopic selection, using machine learning as a guide, can improve the accuracy of photometric redshifts while minimizing the quantity of required spectroscopic observations. Further work needs to be done on optimizing selection of samples before large investments of telescope time on spectroscopy begin.

Classical stratification techniques (e.g., Neyman 1938) exist for an optimum variance estimator for a discrete set of "classes", each with its own variance. However, it is unlikely that such methods will be sufficient given the observational limits on physical quantities observed. Strategies from dimensionality reduction and information gain techniques can help predict which galaxies are most useful for a training set. Sampling strategies using Local Linear Embedding (Roweis & Saul 2000) can preserve the distribution of spectral types in local spectroscopic surveys with around 10,000 galaxies (compared to an initial sample of 170,000). This works by considering how much new information is gained as we add galaxies to the sample (VanderPlas & Connolly 2009). The sample size reduction is consistent with those used for Principal Component Analysis of SDSS Spectra (Yip et al. 2004). Methods based on physical modeling of galaxy spectra are predicted to require even fewer spectra for training (Bordoloi et al. 2012).

***Impact of incorrect training/calibration redshifts*** Using a combination of spectroscopic and photometric simulations, Cunha et al 2012b showed that for a DES-like survey, spectroscopic samples used for calibration of photo-$z$ error distributions must have an incorrect-redshift rate below 1% to avoid significant biases to the dark energy equation of state parameter, $w$; requirements for Stage IV surveys such as LSST will only be more stringent.

This small a failure rate is achieved in only the very highest-confidence classes of redshifts in current spectroscopic samples. Out of the total set of objects with "robust" (nominally >95% secure or better) redshift measurements, existing deep samples have incorrect-redshift rates ranging from 0.3–0.4% (for DEEP2, based on tests with repeated independent measurements) to >3% (for VVDS). Training sets could be restricted to subsets of objects with more secure redshifts, but that would render the set more strongly incomplete (e.g., reducing the redshift success rate from ~75% to ~60% for DEEP2, or to only 21% of galaxy targets for VVDS-wide).

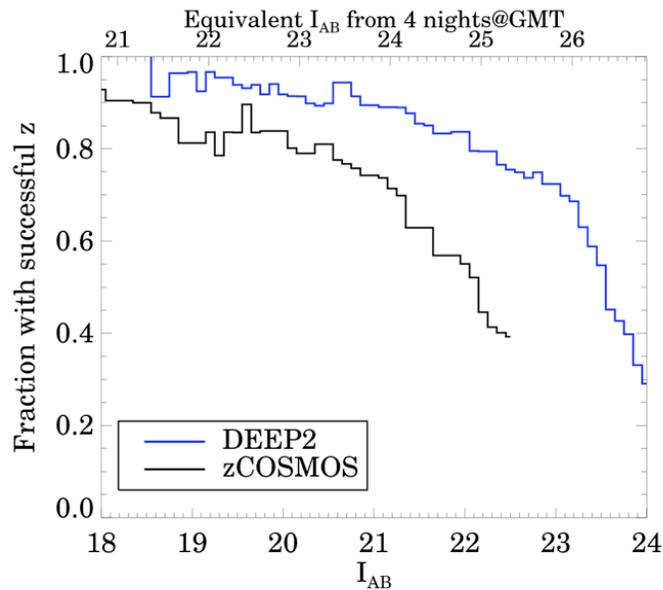

**Figure 2-2.** *Redshift completeness rates from current spectroscopic surveys. The curves show the fraction of objects that yielded a secure (>95% confidence) redshift in the DEEP2 (blue) or zCOSMOS-bright (black) redshift surveys, as a function of i band magnitude ($I_{ab}$). The Dark Energy Survey will use objects to i=23.7 for dark energy, science, while LSST samples will extend to i=25.3; current spectroscopic samples are highly and systematically incomplete even at brighter magnitudes than this. The x axis at top shows the i magnitude that would yield equivalent signal-to-noise with 4 nights of time on the Giant Magellan Telescope (or equivalently, ~100 hours on the Keck telescopes) as DEEP2 or zCOSMOS achieved in one hour of time for the magnitude on the lower x axis. For instance, redshift success rates at the LSST depth of i=25.3 for 4-night GMT observations should be comparable to those in one-hour observations of i=22.5 objects, as the signal-to-noise obtained will be equivalent. The time required to obtain training set samples will be determined by the exposure time to achieve the desired level of completeness, as shown here; the number of spectra that can be observed at once, which can vary from a few dozen to a few thousand; and the area and number of regions of sky to be surveyed.*

Restricting to only the most secure redshifts may still be worthwhile. As Cunha et al 2012b show, the biases in cosmological parameters are a very strong function of the fraction of wrong redshifts in the calibration sample. As shown in Tables 2 and 3 of that work, a decrease of around 2% in the fraction of correct redshifts can imply more than a factor of 5 increase in the bias in the dark energy equation of state parameter, $w$, resulting from photometric redshift calibration errors. This bias has much greater impact than the reduction in statistical errors that would come from a larger (but less secure) sample. The sensitivity of cosmological constraints to incorrect redshifts in a training set suggests that criteria for spectroscopic redshifts have to be heavily weighted towards accuracy if a training set is to be relied upon to determine high-precision calibrations.

The cost of being selective is a much lower redshift success rate, however. The set of objects which yield less-secure redshifts and would ideally be excluded are far from a random subset of all targets in a redshift survey, but rather correspond to those with weaker emission lines and absorption features - i.e., they must have different star formation histories and spectral energy distributions than the set of targets that yield more secure redshifts. As a result of being selective in redshift confidence, more classes of galaxies would be systematically missed in the training set used, or a greater fraction of parameter space would have to be rejected due to incomplete training, increasing uncertainties in redshift distributions and/or reducing sample sizes for dark energy studies.

An alternative approach is to develop photometric redshift algorithms that are robust to the presence of incorrect redshifts in the training set. The development of such robust methods will be of great benefit for future surveys such as LSST. This would not solve the calibration problem, however, unless the false-redshift rates (as a function of all galaxy properties) were extremely well known. We therefore anticipate that unless training sets that are >~99.9% complete and >99% correct over the range of galaxy parameters that will be used for dark energy analyses can be obtained - which has not been demonstrated to date - precision calibration for Stage III and Stage IV surveys will rely on other methods; we discuss promising possibilities in §3.1.

*Impact of sample/cosmic variance:* Poisson errors may not be the dominant source of measurement error in determining redshift distributions (i.e., in photo-z calibration). The number of objects in a spectroscopic sample that are at a given redshift will exhibit additional fluctuations due to the large-scale structure of the Universe itself. Depending on survey characteristics, these fluctuations can dwarf Poisson errors. A particular sight line will pass through regions that are overdense at some redshifts and underdense at others; as a result, the number of galaxies at a given $z$ will fluctuate, a phenomenon generally referred to as sample or cosmic variance. As a further complication, the assembly history of dark matter halos is different in overdense and underdense regions, so they will contain systematically different galaxy populations. This effect can be eliminated only by surveying a large enough number of well-separated regions of the Universe that these statistical fluctuations may be averaged out. The impact of sample variance can be particularly large in training-set-based methods of determining photometric redshifts, as in that case any variance in the spectroscopic training set used will be directly imprinted in redshift distributions everywhere on the sky.

If the spectroscopic samples obtained for training of photometric redshifts are also to be used for calibration (i.e., if we assume that a very high redshift completeness is possible to obtain), the effects of sample variance must be kept to a negligible level. In particular, if our methods are not robust to the effects of sample variance, we must ensure that the **distribution** of spectroscopic redshifts for each photo-$z$ bin (or bin of other observables used for analysis) in the training sample is representative of the true redshift distribution for the full trained sample from an imaging survey (Cunha et al. 2012a). In particular, it is necessary (though not sufficient) that the redshift distribution of the calibration sample closely match that of the true sample (up to smoothly-varying weights that may be incorporated into training algorithms).

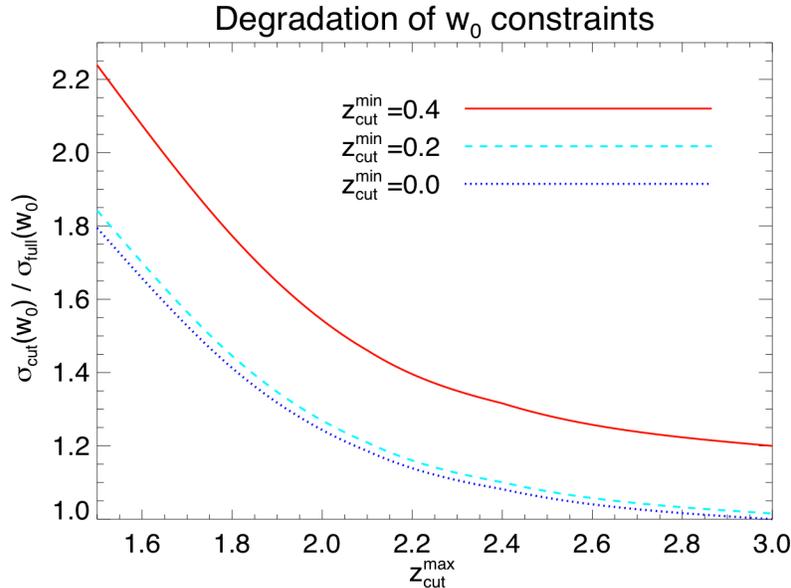

**Figure 2-3.** *The impact on constraints on the current value of the dark energy equation-of-state parameter $w_0$ that arise from excising galaxies with particularly low or high photo-z's from the imaging sample. All curves show errors from an LSST-type survey normalized by the error obtained if the full redshift range of the sample is used. When the imaging data is restricted to the range $0.2 < z_{ph} < 2.2$, the constraining power of the survey is weakened by less than 20%. Such cuts will likely result in a dramatic reduction of the outlier fraction, and come at a comparatively modest cost in the statistical constraining power of future dark energy experiments.*

Unfortunately, most spectroscopic instruments on large telescopes have relatively small fields of view. Because of sample variance, the redshift distribution obtained from any single pointing of these instruments will depart greatly from the true redshift distribution of equivalent samples selected over the entire survey footprint of Stage III and IV imaging experiments. Depending on the number of redshifts obtained and the size of a field, sample variance can be more than an order of magnitude larger than the Poisson noise in redshift distributions (see, e.g., Newman & Davis 2002, van Waerbeke et al. 2006 and Cunha et al 2012a).

Cunha et al 2012a estimate that, for the Dark Energy Survey (5000 deg$^2$, i<23.7) sample, around 100-200 patches, with area between 0.03 and 0.25 deg$^2$ and around 300-400 galaxies per patch, would be needed to ensure that biases due to sample variance in the calibration sample do not dominate the statistical uncertainties in the weak lensing shear-based estimates of the dark energy equation of state parameter, $w$, assuming a direct calibration of redshift distributions via spectroscopic samples. If a significant fraction (> 50%) of the galaxies in a patch are observed, the requirements on the number of patches required can decrease to as few as 60 0.25 deg$^2$ patches. If wider-field spectrographs are used, the sample variance from a single patch will be less, so fewer patches are required. For stage IV surveys, the requirements on sample variance will generally only be greater than for that analysis, as the increase in statistical power implies that any systematics have to be controlled to higher accuracy.

The requirements estimated in Cunha et al 2012a do not take into account a different source of field to field variation: for ground-based imaging experiements, observing conditions will inevitably not be uniform across the sky. These differences may introduce additional spatial variation in the redshift distribution of samples selected based on their photometric properties. Although flux errors being worse in one area than another should not bias photometric redshift estimates in the mean if accounted for properly, the uncer-tainty in photometric redshift must increase in areas with greater uncertainties, yielding broader redshift distributions for samples selected based on any affected property (including $z_{phot}$). As a result, the full photo-$z$ error probability distribution function $P(z_{true}|$observed properties$)$ must be a function of position on the sky if observing conditions vary. This function could be trained in a conditions-dependent way empirically by obtaining spectroscopic samples for regions spanning the range of photometric conditions in the imaging survey; as a higher-dimensional function would need to be constrained, this may make the spectroscopic samples required larger. If the spatial dependence of photometric errors is well understood, however, it should be possible to degrade the photometry in training sets covering good areas to match the errors in worse regions and determine the effects of varying observing conditions without additional spectroscopy. In these regions, galaxy shape measurements may degrade in addition to photometric redshift estimates; this covariance could potentially lead to false shear power spectrum signals.

The sample variance requirements on photometric redshift calibration samples which were derived in Cunha et al. 2012a are based on what could be described as direct, or brute-force calibration, which is the most conservative approach. There are several possibilities for decreasing the requirements, including:
 - Utilizing the observed fluctuations in the number of objects at a given redshift in each field to normalize out the sample variance. As the strength of fluctuations will scale with galaxies' large-scale-structure bias (the strength of their clustering relative to the underlying dark matter), being able to measure the clustering properties of the spectroscopic sample will significantly enhance the power of such corrections.
 - The fluctuations in the redshift distribution of each patch can potentially be estimated through angular cross-correlation analyses, and thereby removed.
 - Bordoloi et al. 2010 developed a procedure to improve individual galaxy redshift likelihoods that seems to decrease requirements on redshift calibration in the context of redshift distribution estimation. However, in the context of cosmological analysis, a full analysis that goes beyond the redshift distribution and involves the full error distribution $P(z_{true}|z_{phot})$ is required to ascertain the success of this method. Analysis of the sensitivity of such techniques to systematic redshift failures within the reduced training set utilized is also necessary.

If we can develop improved methods of taking out sample variance in training sets, both photo-$z$ training and calibration will be more effective. For instance, Cunha et al 2013 (in prep.) show that, if patches are not chosen randomly but instead are picked based on how similar they are to mean of the survey, then requirements decrease by factors of a few, and one can successfully calibrate photo-$z$'s for DES with as few as ~20 patches. The degree of similarity can be established using brighter overlapping spectroscopic samples or even purely photometric properties.

For spectroscopic coverage with a fixed total area, dividing that area amongst more statistically independent fields will always reduce sample variance (cf. Newman & Davis 2002). However, field sizes must be at least several correlation lengths (the typical length scale of the large-scale structure) across to allow accurate measurement of clustering, which is necessary for predicting the strength of sample variance and (potentially) correcting for it. Typical galaxy populations at z~1 have correlation lengths of ~5 $h^{-1}$ Mpc or less (in comoving units; see, e.g., Coil et al. 2008), corresponding to angular scales of 7.4 arcmin or less at z~1, implying that a minimum field diameter of ~20 arcmin is necessary.

Anticipating that at least some degree of optimization and mitigation of sample variance should be possible, for nominal training set samples we assume that a minimum of 15 0.09 deg$^2$ fields (corresponding to areas 20 arcmin in diameter) must be surveyed to obtain spectroscopic datasets that will be well-suited for training and (if redshift failure rates are sufficiently low) for calibration. With this number of training areas, it will be possible to characterize field-to-field variations empirically based on observed scatter. A greater number of fields would improve training and reduce the need for strategies to mitigate sample variance.

## 2.2 Estimates of telescope time required

To summarize the discussion above, photo-$z$ training samples for Stage III and Stage IV surveys will have a number of **minimum** characteristics if they are to serve both training and calibration needs (requirements for training alone may be somewhat smaller, but likely not by much):

*Number of spectra*: Several tens of thousands of spectra will be required at minimum, in order to characterize both core photometric redshift errors and outlier characteristics. Ideal training samples would include >$10^5$ redshifts. For numbers below, we will assume 30,000 spectra as our baseline target for both Stage III and Stage IV surveys.

*Areas and numbers of spectroscopic fields*: Fields should be a minimum of ~20 arcmin in diameter (0.09 deg$^2$ in area) to allow coverage of multiple correlation lengths, allowing mitigation of sample/cosmic variance. A minimum of 10-20 widely-separated fields will be required (with more being desirable); without development of optimized techniques, estimated requirements would reach 50-200 ¼ deg$^2$ fields for DES (Cunha et al 2012a), or ~20 10 deg$^2$ fields (or hundreds of small fields) for LSST (Ma et al. 2011). We will take 15 fields as nominal below.

*Depth and completeness*: We desire a completeness as high as possible while still requiring a feasible total observing time with ground-based telescopes. For DES (with magnitude limit $i$=23.7), we anticipate that 75% completeness spectroscopy would require 5 hours of on-sky time (equivalent to ~7.5 hours total allocated time with weather and overheads) using the DEIMOS instrument at the Keck Observatory, based on the time required to reach equivalent signal-to-noise to current surveys but at the fainter depth of DES. At LSST ($i$=25.3) depth, ~100 hours of on-sky time (=~150 hours with overheads) would be necessary. To achieve 90% completeness, we find that at least six times more exposure time (~30 hours at DES depth or 600 hours for LSST) would be necessary.

For other telescope and instrument combinations, the exposure time required to reach the same signal-to-noise (and hence completeness) will scale as the product of the effective collecting area of the telescope and the total throughput of the instrument; e.g., assuming equal throughput, since the collecting area of the Subaru telescope is 0.7 times as large as that for Keck, the same observations would take 1.43 times longer at Subaru. We will present numbers for both 75% and 90% completeness samples. In general, effective throughputs for yet-unbuilt instruments are only poorly known; we therefore will assume that they match DEIMOS, which may be over-optimistic. We also assume that all spectrographs cover the full optical window (ideally, extending into the near-infrared), from ~0.4 – 1 micron, and that they have sufficient spectral resolution (R=$\lambda/\Delta\lambda$) to resolve the [OII] doublet.

To maximize completeness, we must combine optical spectroscopy from the ground with infrared spectroscopy in space. Here, we provide estimates only of required ground campaigns, as those may begin sooner and their completeness is better-understood currently. Redshift success may be somewhat higher than 75%/90% (numbers based on experience with DEEP2 and DEEP3) if high spectral resolution is obtained to longer wavelengths (e.g., to 1.3 microns for Subaru / PFS) and throughputs are high; nonetheless, based on past experience we do not anticipate radical improvements from modest increases in spectral range, especially for ground-based samples where night sky emission lines dominate the near-infrared. On the other hand, the actual yield may turn out to be lower than 75%/90% if the definition of success must be limited to extremely (>99%) secure redshifts for calibration purposes, as described above. We may hope that the combination of ground and space-based spectroscopy may yield much lower failure rates, approaching 1% or less, but that has not been demonstrated to date. If samples are incomplete, less direct methods of calibration will be needed; see §3.1 for details. Incomplete spectroscopic samples will nevertheless be highly useful for training of photometric redshift algorithms, especially for template-based methods.

*Multiplexing*: Photometric redshift training samples will generally be infeasible to obtain one object at a time, due to the large exposure times required. Instead, we will rely on multi-object spectrographs that may obtain many spectra at once (up to 5,000 in the case of the DESI instrument).

*Basic scalings*: Depending on instrument characteristics, spectroscopic samples will be *fields-limited* (if multiplexing and the field of view of the spectrograph are both large, simply reaching ∼15 fields will determine the required total number of pointings); *FOV-limited* (if multiplexing is high but field of view is low, more pointings will be required to cover ∼15 x 0.1 deg$^2$ total sky area); or *multiplex-limited* (if field of view is large but multiplexing is low, many pointings will be required to get spectra of ∼30,000 targets). The exposure time required will be (assuming 1 night = 8 hours' time):

*Fields-limited*: $t_{observe}$ =(280 nights) × (Nfields / 15) × (Equivalent Keck/DEIMOS exposure time / 100 hours) × (0.67 / observing efficiency) × ( 76 m$^2$ / telescope effective collecting area)× (0.3 / [telescope + instrument throughput])

*FOV-limited*: $t_{observe}$ =(280 nights) × (Nfields / 15) × (314 arcmin$^2$ / area per pointing )×( Equivalent Keck/DEIMOS exposure time / 100 hours) × (0.67 / observing efficiency) × ( 76 m$^2$ / telescope effective collecting area) × (0.3 / [telescope + instrument throughput])

*Multiplex-limited*: $t_{observe}$ =(280 nights) × (Nobjects / 3×10$^4$) × (2000 / Number of simultaneous spectra)×(Equivalent Keck/DEIMOS exposure time / 100 hours) × (0.67 / observing efficiency) × ( 76 m$^2$ / telescope effective collecting area) × (0.3 / [telescope + instrument throughput])

For any given telescope/instrument combination, the actual exposure time required will be the **greatest** out of the fields-limited, FOV-limited, and multiplex-limited values.

*Candidate telescopes and instruments*: For this white paper, we do not consider every possible telescope and instrument combination, but rather the most relevant instruments expected to be available in the 2020-2030 timeframe that bracket the range of possibilities. We restrict the list to instruments capable of resolving the [OII] doublet (spectral resolution $\lambda/\Delta\lambda$>∼3500) over a broad spectral range (∼0.5–1μm at minimum), as that is key for providing highly-secure redshifts for galaxies at 1<$z$<2, a critical range for dark energy studies.

Other telescopes or instruments may provide similar characteristics; e.g., 4MOST (de Jong et al. 2012) would be broadly similar to DESI (Schlegel et al. 2011), while the proposed Next Generation CFHT (ngCFHT) and Southern Spectroscopic Survey Telescope (SSST) would have roughly similar characteristics to PFS on Subaru (Sugai et al. 2012). We also consider the workhorse DEIMOS spectrograph on the Keck telescope (Faber et al. 2002) as a baseline example of what can be done with current technologies. Of the items listed in the table, Keck/DEIMOS is available now; PFS, VLT/MOONS (Cirasuolo et al. 2012), WHT/WEAVE (Dalton et al. 2012), and DESI should become available around 2018; while GMT/MANIFEST+GMACS (Saunders et al. 2010, DePoy et al. 2012), TMT/WFOS (Pazder et al. 2006), E-ELT/OPTIMOS (Hammer et al. 2010, Le Fevre et al. 2010), SSST, and ngCFHT would be available around 2022 at earliest.

| Telescope / Instrument | Collecting Area (m²) | Field area (arcmin²) | Multiplex | Limiting factor |
|---|---|---|---|---|
| Keck / DEIMOS | 76 | 54.25 | 150 | Multiplexing |
| VLT / MOONS | 58 | 500 | 500 | Multiplexing |
| Subaru / PFS | 53 | 4800 | 2400 | # of fields |
| Mayall 4m / DESI | 11.4 | 25500 | 5000 | # of fields |
| WHT / WEAVE | 13 | 11300 | 1000 | Multiplexing |
| GMT/MANIFEST+GMACS | 368 | 314 | 420-760 | Multiplexing |
| TMT / WFOS | 655 | 40 | 100 | Multiplexing |
| E-ELT / OPTIMOS | 978 | 39-46 | 160-240 | Multiplexing |

**Table 2-1.** *Characteristics of current and anticipated telescope/instrument combinations relevant for obtaining photometric redshift training samples. Assuming that we wish for a survey of ~15 fields of at least 0.09 deg² each yielding a total of at least 30,000 spectra, we also list what the limiting factor that will determine total observation time is for each combination: the multiplexing (number of spectra observed simultaneously); the total number of fields to be surveyed; or the field of view of the selected instrument. For GMT/MANIFEST+GMACS and VLT/OPTIMOS, a number of design decisions have not yet been finalized, so a range based on scenarios currently being considered is given.*

| Telescope / Instrument | Total time(y), DES / 75% complete | Total time(y), LSST / 75% complete | Total time(y), DES / 90% complete | Total time(y), LSST / 90% complete |
|---|---|---|---|---|
| Keck / DEIMOS | 0.51 | 10.22 | 3.19 | 63.89 |
| VLT / MOONS | 0.20 | 4.00 | 1.25 | 25.03 |
| Subaru / PFS | 0.05 | 1.10 | 0.34 | 6.87 |
| Mayall 4m / DESI | 0.26 | 5.11 | 1.60 | 31.95 |
| WHT / WEAVE | 0.45 | 8.96 | 2.80 | 56.03 |
| GMT/MANIFEST+GMACS | 0.02 - 0.04 | 0.42 - 0.75 | 0.13 - 0.24 | 2.60 - 4.71 |
| TMT / WFOS | 0.09 | 1.78 | 0.56 | 11.12 |
| E-ELT / OPTIMOS | 0.02 - 0.04 | 0.50 - 0.74 | 0.16 – 0.23 | 3.10 - 4.65 |

**Table 2-2.** *Estimates of required total survey time for a variety of current and anticipated telescope/instrument combinations relevant for obtaining photometric redshift training samples. Calculations assume that we wish for a survey of ~15 fields of at least 0.09 deg² each, yielding a total of at least 30,000 spectra. Survey time depends on both the desired depth (i=23.7 for DES, i=25.3 for LSST) and completeness (75% and 90% are considered here). Exposure times are estimated by requiring equivalent signal-to-noise to 1-hour Keck/DEIMOS spectroscopy at i~22.5. GMT / MANIFEST + GMACS estimates assume that the full optical window may be covered simultaneously at sufficiently high spectral resolution; in some design scenarios currently being considered, that would not be the case, increasing required time accordingly.*

*Options for near-infrared spectroscopy* Near-infrared spectroscopy is needed to ensure that multiple emission lines and/or absorption features are observable for galaxies over the full redshift range of interest, which is prerequisite for secure redshift determinations. The coverage of optical spectrographs available today typically reaches 9000-9500 Å, which implies that the [OII] emission-line doublet (the bluest of the strongest emission lines) can be identified up to redshifts around z=1.4-1.5 at most. Similarly, the most significant absorption feature, the 4000 Å break, can be identified out to redshift ~1.3. The proposed Prime Focus Spectrograph would have coverage at up to ~1.3 microns, enabling good coverage at redshifts up to $z$~2.5 (assuming very long exposure times) but not beyond.

With only optical spectroscopy, it is still possible to derive spectroscopic redshifts beyond those limits, but at a significantly lower confidence level and generally in a less complete manner. For instance, rest-ultraviolet absorption lines from Fe and Mg can provide redshifts for objects at $z$~2, but these lines are comparatively weak (requiring high signal-to-noise) and only present in strongly star-forming stellar populations, so will yield redshifts for only limited subsets. Based on the COSMOS Mock Catalogs (Jouvel et al. 2009), we estimate that about 10% of DES objects (i<23.7) and 23% of the LSST Gold Sample (i<25.3) should be at z>1.4. In order to utilize the full redshift range of these experiments for dark energy analyses, we will require training sets that extend to high redshift, which in turn will require spectroscopy in the near-infrared.

However, due to the brightness of the night sky at infrared wavelengths, deep spectroscopy in the near-infrared is extremely time-expensive to pursue from the ground. The most powerful near-IR multi-object spectrographs currently available are KMOS (K-Band Multiobject Spectrograph) at VLT and MOSFIRE at Keck. For a survey such as DES, about 10% of the photometric sample is above z~1.3, which is roughly the region beyond which optical spectroscopy has a severe drop in efficiency. Since complete spectroscopy of the order of 20,000-30,000 galaxies would be needed to calibrate DES total, we might expect that a minimum of 2-3,000 redshifts would have to be obtained using IR spectroscopy. One possibility may be to cull potentially high-redshift objects from dark energy samples using machine learning techniques (see, e.g. Cunha et al 2012a,b; though the tendency of objects at high redshifts to have flat spectral energy distributions, which yield similar colors at all redshifts, may make that difficult). In that case, one would still require a complete sample of several thousands of galaxies, and hence at minimum several hundreds to a thousand galaxies above z~1.3, to reasonably map the location of the high-z galaxies in color space. KMOS can obtain spectra for up to 24 objects within a 7.2 arcmin diameter (0.01 deg$^2$) field of view, while MOSFIRE can obtain up to 46 simultaneous spectra over a 6.1 arcmin x 6.1 arcmin (0.01 deg$^2$) area; hence, obtaining sufficient spectra over nonnegligible areas while sampling many fields would require a large time investment with these instruments.

For a survey such as LSST, discarding objects at $z$>1.4 has greater effect, as the galaxy redshift distribution has a long tail extending to z~4. In addition, the integration time required for high redshift completeness to LSST depth is prohibitive. Simulations find that with very large exposure times, of the order of 50 hours, secure redshifts for close to 90% of the targets with LSST Gold Sample depth can be obtained using the PFS spectrograph (Cunha et al. 2013). This is somewhat more optimistic than extrapolations from current surveys for constant signal-to-noise would indicate, possibly due to inaccuracies in the COSMOS mock catalog (Jouvel et al. 2009) used here. The resulting sample would be highly complete up to z~1.8. Ground-based infrared spectrographs could improve the situation at higher redshifts, but only for limited samples over small areas.

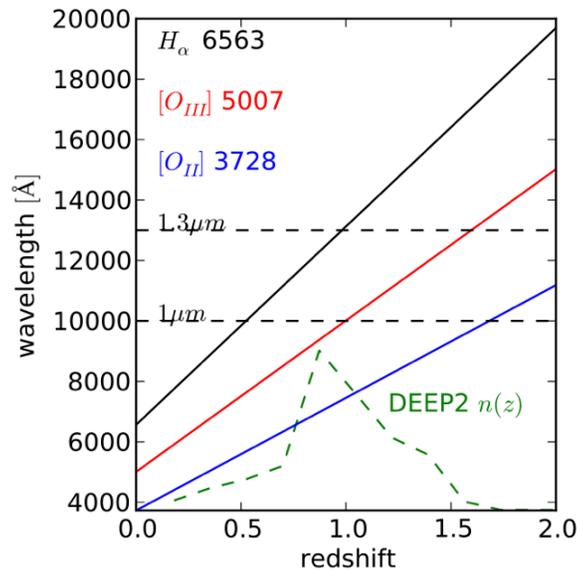

**Figure 2-4.** *Illustration of the need for infrared coverage in spectroscopic samples, drawn from Comparat et al. 2013 (submitted). Black, red, and blue curves show the observed wavelengths of the Hα, [OIII], and [OII] spectral features as a function of wavelength. At spectral resolution ($R=\lambda/\Delta\lambda$) of a few hundred, [NII] and [SII] lines may be detected near Hα if they are strong enough, the [OIII] doublet will resolve into two spectral lines (4959 and 5007 Å), and the Hβ (4861 Å) line will be similar in wavelength to [OIII]; depending on galaxy properties [OIII] may be strong and Hβ weak or vice versa. At higher redshifts, the [OII] 3727 Å doublet may be the only feature in the optical window; a minimum resolution of R~4000 is required to split the two doublet lines and obtain a secure redshift measurement. For quiescent galaxies without star formation or nuclear accretion activity, the most readily available spectral features are the Ca H & K absorption lines and the 4000 Å spectral break, which are similar in wavelength to [OII]. A secure redshift requires significant (>~ 5 sigma) detections of multiple spectral features, as that will unambiguously determine the z. At redshifts beyond z=1.5, the [OII] doublet passes beyond the optical range ($0.4 < \lambda < 1$ micron), and infrared spectroscopy is in general necessary to obtain secure redshifts (rest-ultraviolet absorption features can provide redshifts for highly star-forming galaxies at higher z, but these features are relatively weak and require high sensitivity at blue wavelengths).*

In contrast, observations from space are minimally affected by emission from the Earth's atmosphere, so can detect faint spectral features in much less time than from the ground. Both the *Euclid* and *WFIRST-2.4* satellites are designed to obtain low-spectral-resolution, relatively shallow (particularly in the case of *Euclid*) infrared spectroscopy using grisms in order to detect the baryon acoustic oscillation feature in the clustering of galaxies at $z>1.5$. The Euclid design includes a slitless spectrometer with $R=\lambda/\Delta\lambda \sim 250$ that should obtain millions of spectra (primarily from Hα at $0.7<z<2.0$). WFIRST, with a ~0.3 deg$^2$ grism and R~500-800, will obtain ~20 million emission line galaxy redshifts in the 2000 deg$^2$ High Latitude survey covering $1<z<3$. Although these experiments are focused on brighter objects, some objects which fail to yield redshifts in ground-based spectroscopy will provide multiple sufficiently strong emission features within the wavelength windows covered by these satellites, even given their low spectral resolution compared to ground-based optical spectroscopy (which will cause some emission lines to be blended together and hence undetectable). The surveys will by necessity cover large areas, so sample/cosmic variance will be of negligible concern.

A particularly interesting capability would be provided by the planned IFU spectrograph for *WFIRST*-2.4 (Spergel et al. 2013a,b); due to lower backgrounds, it can provide significantly deeper spectroscopy than grism spectroscopy. The proposed instrument concept covers a sufficiently large area of sky that an LSST gold sample object should be on the IFU 50% of the time. If operated in parallel to WFIRST imaging in a blind mode, the IFU can provide more than 40,000 relatively deep infrared spectra, which can provide high redshift completeness for objects at redshifts up to $z$=3.7 (Cunha et al. 2013); deep exposures on supernovae can yield higher signal-to-noise. Because these spectra would be scattered over the full WFIRST imaging footprint, sample/cosmic variance would be a negligible issue. Finally, WFIRST includes plans for both large and small guest observer programs (1.4 years of the 5 year mission), providing an opportunity for proposals for targeted spectroscopic and near-infrared follow up of LSST targets to fill gaps in training data.

Ultimately, the most sensitive instrument for obtaining secure redshifts for objects that do not yield multiple features from ground-based spectroscopy would be the NIRSPEC spectrograph for the *James Webb Space Telescope*. Due to the much larger collecting area, it will be far more sensitive than other options. NIRSPEC should be capable of delivering intermediate-resolution ($R$~1000, capable of resolving most major emission lines but not the separate components of the [OII] doublet) for roughly 100 objects simultaneously over a 3 arcmin x 3 arcmin (0.003 square degree) area, if its microshutter arrays do not suffer further failures on launch (in which case it could still provide spectroscopy for one object at a time). Because of the limited field of view, sample/cosmic variance will be the biggest limiting factor for JWST spectroscopy; large allocations of time would be required to cover large numbers of ~0.09 deg² fields, though required exposure time per pointing to reach LSST depths should be relatively short compared to other options in the near-infrared.

Space-based infrared spectroscopy should increase redshift completeness to substantially greater levels than that achieved by ground-based data alone. Whether it is capable of delivering (in concert with deep optical spectroscopy) the 99-99.9% completeness required for training sets to be used for calibration purposes is still unknown, however, largely as future capabilities will significantly exceed those currently available.

*Calibration and Training of Photometric Redshifts for Supernovae and Galaxy Clusters*
Large imaging surveys such as LSST will discover a million type Ia supernovae (SNIa). These immense SNIa samples hold the promise to make very precise and determinations of cosmological distances in a manner independent of most other probes. Accurate measurements of both axes of the Hubble diagram (redshift and distance) are crucial for any SNIa cosmology survey. A well-calibrated densely sampled network of distance measurements from LSST SNIa would make substantial contributions in determining the nature of dark energy (LSST Science Book; Ivezic et al. 2009).

Redshifts are traditionally measured spectroscopically from either narrow galaxy lines or broad SN absorption features. However, to make use of the full range of SNIa from future imaging surveys, we will need to develop the capability to determine the type and redshift of supernovae based primarily on multi-band photometric data - another application of the idea of photometric redshifts. With ~10,000 SNIa in each $\Delta z$=0.05 wide bin in redshift, statistical errors in the distance to that bin will be ~100x smaller than the uncertainty in the distance to a single supernova. As a result, systematic errors should dominate, and the *calibration* of photo-z's for SNIa is a key need.

In the ideal limit of a perfectly-understood redshift distribution, SNIa from the LSST sample have the potential to measure distances with errors of approximately 0.1% in bins of $\Delta z$=0.05 from 0.1<z<0.8, and thereby constrain the dark energy equation of state parameter at the level of $\sigma(w_{piv})$ ~ 0.02. To approach this goal, the photometric redshift distribution of SNIa will need to be trained and calibrated in a manner similar to the galaxy photometric redshift distribution (Barris et al. 2004; Wang 2007; Wang et al. 2007;

Zentner & Bhattacharya 2009; Asztalos et al. 2010; Kessler et al. 2010; Gong, Cooray & Chen 2010; Kim & Linder 2011; Sako et al. 2011). The critical need to obtaining well-calibrated photometric redshifts for SNIa samples is to have a representative sample of all types of supernovae from the apparent brightness range of the SNIa sample.

Using the same formalism as described in §2.1, it is estimated that constraints on $w_{piv}$ within a factor of two of the ideal value can be achieved with calibration samples of between ∼2000 and ∼6000 spectroscopic redshifts for SNIa out to redshift $z$∼1 (Zentner & Bhattacharya 2009). This leads to the conclusion that the number of spectroscopic redshifts of active supernovae required for photometric redshift calibration should be comparable to the total number of supernovae in contemporary samples; such a sample should be feasible to obtain with the instruments available in the next two decades. Unlike the case for galaxies, supernovae exhibit strong spectral features, making the problems of incompleteness and incorrect redshifts that affect photo-z calibration for galaxy samples negligible for supernovae.

However, complex degeneracies between reddening, redshift, and intrinsic SN colors limit the current state-of-the art photometric redshift analysis techniques and lead to redshift biases at the moment that are prohibitively large (Kessler et al. 2010; Sako et al. 2011). Without a breakthrough in analysis methods and improved understanding of the astrophysics of supernovae and their host galaxy environments, it is quite possible that LSST-type surveys must obtain spectroscopic redshifts of the SN host galaxies to construct a useful Hubble diagram. These will be large samples of hundreds of thousands of galaxies.

Those samples may not be overly expensive to obtain. Obtaining spectroscopy for all of the host galaxies of supernovae detected in the deep-drilling regions of large imaging surveys (which will provide the bulk of the supernova-based constraining power from these projects) can be efficiently accomplished with wide-field multi-object spectrographs on 4-m and 8-m telescopes. A DESI-like instrument could obtain ∼75%-complete samples of host galaxy redshifts for the ∼300,000 supernovae found in LSST deep-drilling fields with ∼60 clear nights of telescope time (Schlegel et al. 2011). This new spectroscopy would have to lag the supernova observations themselves in time, but could enable substantially stronger dark energy constraints. Unlike for other probes of dark energy, incompleteness in this spectroscopic dataset would have limited impact, increasing random errors in the resulting sample by ∼15% but with little, if any, systematic effect.

Such spectroscopy would have impact beyond supernova-based dark energy studies. It could contribute to photometric redshift training sets at the bright end and to samples for cross-correlation studies (q.v. below). In addition, these redshift measurements will enable studies of the rate of supernovae as a function of SN type, galaxy properties, and redshift, which provides key constraints on models of supernova progenitors (which are still poorly understood). Spectroscopy of supernovae in optically-faint hosts could also improve photometric redshift training for galaxy samples, as some supernovae will be much brighter than the galaxies they reside in, enabling redshift measurements for otherwise-impossible galaxies.

Similarly, specialized spectroscopy may enhance the use of galaxy clusters for cosmology. For such studies, it will be important to understand how well photo-z codes trained on field populations describe faint cluster members, as well as other objects along cluster lines of sight. These questions may be addressed through deep spectroscopy of fields containing galaxy clusters spanning a range of redshift and mass. Ideally, this would entail comprehensive spectroscopy of objects in ∼20 cluster fields, with ∼500-1000 objects per field, to similar depth as for training/calibration datasets described in section 2.2. However, required field sizes are somewhat smaller, given the angular size of clusters, making large-aperture (30m+), comparatively small-field-of-view telescopes ideal for this work.

## 3.1 Spectroscopic requirements for cross-correlation methods

As discussed in the preceding section, it is possible that the spectroscopic training sets obtained for Stage III and Stage IV dark energy imaging experiments will not be sufficiently complete to use for a final, high-precision calibration. However, even if our photometric redshift measurements have systematic biases (e.g., due to incompleteness in the training set), so long as we can determine the level of those biases, dark energy inference will generally remain unscathed. A variety of cross-correlation techniques may be used to determine the actual redshift distribution of a sample, allowing us to detect biases that the likely incomplete spectroscopic training sets cannot. Hence, **cross-correlation techniques present the most likely solution for precision calibration of photometric redshifts for dark energy experiments**. However, they are difficult to use for training photometric redshift algorihtms, so deep training sets will still be required for ongoing and future imaging experiments to reach their full potential.

In this section, we consider cross-correlations solely between galaxy positions in imaging and spectroscopic samples; however, other cross-correlations (e.g., with lensing signals) can add additional information not included here. More details on these methods are provided in the CF5 white paper on cross-correlations (Rhodes et al. 2013).

Measuring the cross-correlation between the locations on the sky of galaxies of known spectroscopic redshift with a photo-$z$-selected sample, as a function of spectroscopic $z$ in the former sample, provides sufficient information to reconstruct the true, underlying redshift distribution of the photo-$z$-selected sample. This method can achieve high accuracy even if it is only possible to obtain spectroscopic redshifts for a bright, biased subset of the galaxies at a given redshift (Newman 2008). Cross-correlation techniques rely on the fact that all galaxies -- from the faintest to the brightest -- cluster together on large scales, tracing the underlying web of dark matter. Hence, galaxies in a spectroscopic sample that are at a given redshift will cluster on the sky solely with the subset of galaxies in other samples that are at or near that redshift. We can exploit this fact to accurately determine the true redshift distribution of objects in a photometrically selected sample based on the amount of clustering with galaxies that we know to be at a given redshift.

If we only measure cross-correlations, redshift distribution estimates are degenerate with the strength of the intrinsic, three-dimensional clustering between the two samples; however, the autocorrelations (i.e., degree of clustering with itself) of the photometric and spectroscopic samples -- some of the most basic measurements that are generally made from galaxy surveys -- can break the degeneracy between redshift distributions and clustering in the mean, though clustering evolution remains a factor (discussed in more detail below). Other cross-correlation techniques for testing photometric redshifts have been developed (Zhan & Knox 2006, Schneider et al. 2006) and are likely to be very useful for future dark energy experiments, but they do not break that degeneracy. Further advances on all these current methods (e.g., utilizing lensing tomography as a consistency check, by applying the techniques from Clowe et al. 2013 [in prep.]) could yield even better results.
Again, we refer the reader to the Snowmass white paper on Cross-Correlations (Rhodes et al. 2013) for more details; here, we wish to focus on the requirements for spectroscopic samples to be used in cross-correlation analyses.

Newman (2008) found that, based on conservative estimates of errors from cross-correlation methods, calibrations of sufficient accuracy for LSST could be obtained with spectroscopy of ~25,000 objects per unit redshift (with only a few thousand per unit $z$ required at $z$>2). However, unlike the case for photo-$z$ training, **the spectroscopic sample need not resemble the photo-z objects in properties, but only overlap in redshift and sky coverage**. This allows systematically incomplete surveys that target only the brightest objects at a given $z$ to be used for calibration. Matthews & Newman (2010) showed that the estimates of Newman 2008 are appropriate only for surveys that cover a wide area (a minimum of hundreds of square degrees of overlap between spectroscopic and photometric samples); however, McQuinn and White (2012) demonstrated that, if optimal estimators are used for the analysis, errors a factor of 3 to 10 smaller than those found by Newman 2008 are obtained in this modestly-wide-area limit.

As a result, for cross-correlation calibration of Stage IV surveys we require spectroscopy of a minimum of $\sim 5 \times 10^4$ objects over hundreds of square degrees, with coverage of the full redshift range of the objects whose photometric redshifts will be calibrated; requirements for Stage III surveys will be somewhat weaker. Such a sample could be provided by overlap between dark energy imaging experiments and proposed baryon acoustic oscillation experiments. Those projects yield optimal results when targeting the brightest galaxies at a given redshift over the broadest redshift range and total sky area possible; all of these desiderata also improve the strength of cross-correlation measurements. As with other calibration techniques, all redshifts used must be highly secure (>99+% probability of being correct); however, incompleteness is not an issue.

The eBOSS baryon acoustic oscillation experiment should obtain redshifts of ~1.5 million galaxies and QSOs over the redshift range $0.6 < z < 4$ over 7500 square degrees of sky (Comparat et al. 2013). It will have a minimum of 500 square degrees of overlap with DES, and likely more with HSC, LSST, WFIRST-2.4, and Euclid. Objects at $z<0.6$ from the SDSS and BOSS surveys provide lower-redshift coverage over the same sky area. Hence, eBOSS (in combination with previous Sloan surveys) will provide at least 100,000 redshifts within the footprints of all of the upcoming dark energy experiments. **The eBOSS survey can potentially provide a sufficiently accurate calibration that random, not systematic, errors should dominate in dark energy analyses.** The greatest source of uncertainty is whether the subset of objects that provides highly-secure redshifts in eBOSS will still be sufficiently large for calibration purposes. The proposed Sumire project to be undertaken with the PFS spectrograph on Subaru would have similar impact on dark energy studies (de Putter et al. 2013).

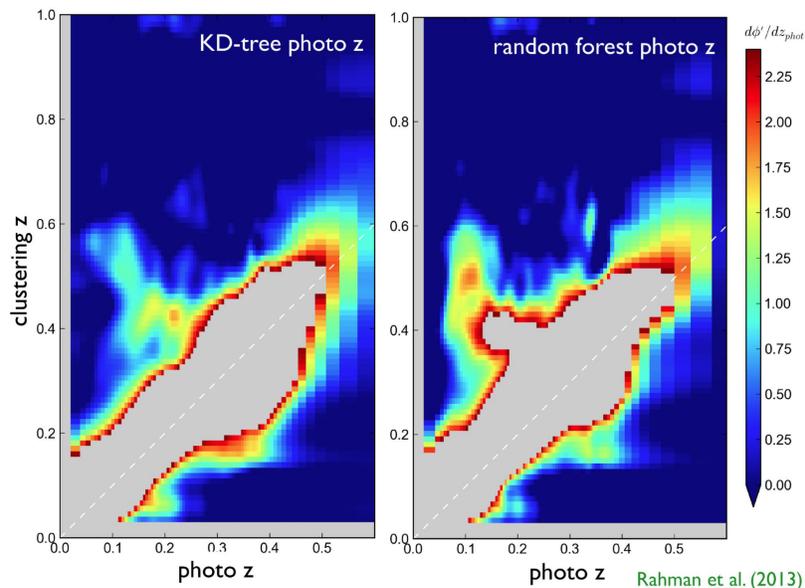

**Figure 3-1.** *Application of cross-correlation reconstruction techniques to the problem of testing photometric redshift algorithms. In this figure (from Rahman et al. 2013 [in prep.]), photometric redshifts derived via two different algorithms were tested via cross-correlation (labelled as 'clustering z' here) with a broad range of spectroscopic datasets in a succession of narrow bins in photo-z. Although the two photometric redshift algorithms perform similarly in the core of the error distribution (truncated by the color scale here), they have significantly different outlier characteristics, readily detectable by this method. This work assumes a constant bias between the clustering of galaxies and dark matter for each photo-z bin; that assumption must be improved upon for application to future dark energy experiments.*

The MS-DESI baryon acoustic oscillation experiment should obtain redshifts of ~20 million galaxies and QSOs over the redshift range $0.6 < z < 4$ over more than 14,000 square degrees of sky (J. Newman, priv. comm.). Because of the wider sky coverage, it is likely to have thousands of square degrees of overlap with DES, and will have more with LSST, *WFIRST-2.4*, and *Euclid*. Again, the SDSS and BOSS surveys will provide lower-redshift coverage over the same sky area; absorption-line systems detected in MS-DESI QSO spectra can provide even more redshifts at low $z$ (Ménard et al. 2013). We anticipate that MS-DESI will provide at least 1,000,000 redshifts within the footprints of all of the upcoming dark energy experiments. **MS-DESI will allow detailed analyses of photometric redshift calibration, rendering calibration errors negligible and providing multiple cross-checks for systematics** (e.g., by comparing redshift distributions reconstructed using tracers of different clustering properties, or restricting to only the most extremely secure redshifts).

*Potential systematics* Work remains to be done to ensure that cross-correlation methods can fully meet the calibration needs for future dark energy experiments. Fundamentally, cross-correlation methods measure a product of the bias (the strength of clustering of a given population to the clustering of the underlying dark matter), $b(z)$, and the probability an object in a photometric sample is actually at a given redshift, $p(z)$. The bias is observed to be a function of both galaxy properties and redshift; the fractional evolution in bias within any photometric sample, $(db/dz / <b(z)>)$, must be known to ~10-50% for biases in $<z>$ reconstruction to not contribute significantly to dark energy uncertainties.

A variety of options for characterizing $db/dz$ exist. In simulations, the ansatz that the bias evolution for photometric samples is proportional to the bias evolution of the spectroscopic sample used for cross-correlation (which is directly observable), works well (Matthews & Newman 2010). Such a model can be tested or improved by doing cross-correlation reconstruction for a series of overlapping bins in $z$ and/or color, since a measurement of the function $b(z) p(z)$, together with the fact that $p(z)$ is a probability distribution function (and hence has fixed integral), provides a measurement of the object-weighted mean of $b$. Therefore, $<b>(<z>)$ is measurable, and may be used to constrain $b(z)$. Gravitational lensing of the cosmic microwave background will provide another constraint on the mean bias of a sample with a different redshift weighting, and hence will also help to constrain $db/dz$. Finally, the bias as a function of restframe color and luminosity is observed to be a smooth function, though one that evolves with redshift, at both $z$~0 and $z$~1 (Cooper et al. 2006, Coil et al. 2008, Zehavi et al. 2011); one can fit this function based on galaxies that do yield redshifts. If the function $b$(restframe color, luminosity, $z$) is known (again, at least to ~10-50%), this is sufficient to predict $b(z)$ for a given galaxy sample, since the observed magnitudes of each object in the sample provide its restframe color and luminosity contingent on being at a given $z$ (the process of k-correction, cf. Blanton & Roweis 2007).

Nevertheless, although these avenues exist, they have not yet been tested at the accuracy required for future dark energy experiments. The greatest concerns would be if objects that have unusual colors (and hence exhibit catastrophic photometric redshift errors) also have unusual clustering for their restframe color (see, e.g., Bernstein & Huterer 2010). New large redshift samples from the BOSS survey provide the best avenue for a clean test of these methods; redshift information from subsets of the sample selected to have different color (and hence bias) from the remainder can be blinded and then reconstructed using the remainder of the sample.

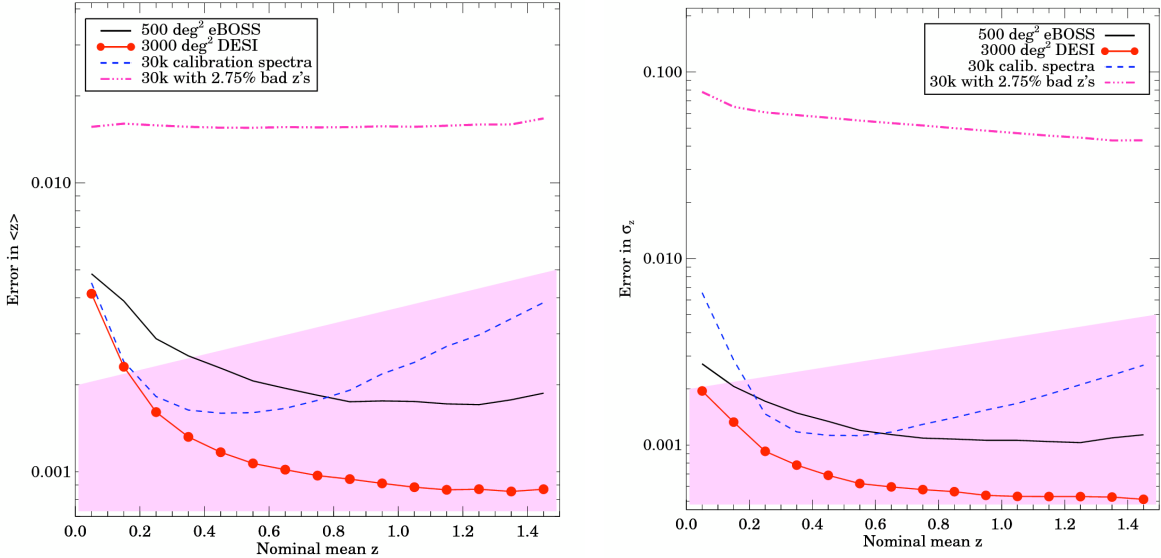

**Figure 3-2.** *Predicted calibration errors for LSST samples in $\Delta z=0.1$ bins resulting from either cross-correlation techniques (discussed in §3.1) or targeted training sets (as discussed in §§2.1-2.2). The plots show the errors in reconstructing the mean (left) or standard deviation (right) of the redshift distribution of LSST photo-z selected samples of different mean redshift (with assumed $\sigma_z=0.05(1+z)$ errors) resulting from cross-correlation (solid) or training-set-based calibrations (dashed). For cross-correlation, we assume an eBOSS sample with an overlap area of 500 square degrees (black line) or a DESI sample with an overlap area of 3000 square degrees (red line). Results may be compared to the results of calibration via an ideal, 100% complete, 100% correct, zero-cosmic variance training set of 30,000 spectra matching the full LSST sample's properties (blue line); such a dataset is unlikely to be feasible to obtain with current or near-future facilities (existing spectroscopic samples to Stage III imaging depth are ~ 50 − 75% complete, >95-99% correct for nominally successful redshift measurements, consist of a few thousand spectra total resulting from dozens of nights on 8-10m telescopes, and have substantial cosmic variance). The difference is stark when the 2.75% erroneous-redshift rate obtained in some deep redshift samples such as VVDS or zCOSMOS is taken into account (purple dot-dashed line). The pink shaded region shows the estimated requirements on our knowledge of these parameters for LSST (errors below $0.002(1+z)$. These simulations do not utilize the optimal estimator of McQuinn & White (2013), which is predicted to yield errors a factor of three to ten lower than the model used here for cross-correlation methods. Even without such improvements, cross-correlation techniques are sufficient to achieve the necessary calibrations for Stage III and Stage IV surveys except at the very lowest redshifts, which can be excised with minimal impact on dark energy inference due to the small number of objects in those redshift bins.*

A second potential source of systematic errors is the cross-correlations induced by gravitational lensing, if not corrected for properly. For flux-limited samples, objects in the foreground will tend to correlate in position with objects in the background due to fainter background objects that are magnified and boosted over the limit. This effect will cause cross-correlations between objects in the spectroscopic sample at a given $z$ and photometric objects that are either in front or behind. The strength of this effect is predictable from the redshift distributions and the strength of biasing of each sample, combined with the flux distribution of galaxies near the nominal survey limit (which may be determined empirically using the deeper subregions included in most imaging-based dark energy experiments). This effect must be accounted for in fitting redshift distributions to achieve the necessary accuracy for photometric redshift calibration. In principle, lensing-induced cross-correlation provides an additional observable that could be used to strengthen bias constraints and hence improve cross-correlation reconstruction, rather than degrading it, but that has not yet been tested.

## 4.1 Important areas for new research

*Using multi-wavelength data to improve calibrations* As stated in Section 2.2, we would like a representative training set of a minimum of ∼30,000 spectroscopic galaxies complete to LSST depths. Spectroscopy alone may not be a realistic solution, given both the telescope time and financial resources necessary, as well as the fact that faint galaxy surveys fail to obtain redshifts for a significant fraction of the galaxies that they observe (see Section 2.1). One approach to address these problems is to supplement existing photometric data for galaxies that fail to yield spectroscopic redshifts with measurements in additional photometric bands to calculate "super-photo-$z$'s". Adding bands both blueward in the UV and redward in the infrared, and potentially narrower bands in the optical, can improve photometric redshift estimation greatly. The photo-$z$ uncertainties and outlier rates for such samples are considerably reduced, as they can detect multiple features such as the Lyman and Balmer breaks simultaneously, breaking degeneracies and constraining redshift solutions. These super-photo-$z$'s can then be used as secondary calibrators for the main photo-$z$ sample. A training set of spectroscopic galaxies is still necessary to calibrate the super-photo-$z$'s, but the selection and sampling can be better tuned for efficient training.

The J-PAS survey (Cenarro et al. 2012) intends to employ such super-photo-z's to perform baryon acoustic oscillation measurements, obtaining higher-precision photometric redshifts for objects to magnitude ∼22.5 (significantly brighter than the DES and LSST survey limits). MKIDS technology could enable an alternative method with similar impact. The proposed Giga-z camera (Marsden et al. 2013) would effectively obtain 30-band imaging of a given region of sky simultaneously, rather than via successive application of different filters. Such imaging would enable high-precision photometric redshifts, which could be used for training broad-filter photo-z's, but still would require spectroscopic training sets to tune the many-band photo-z's, just like the super-photo-z's described above.

The potential synergies between ground-based and upcoming space-based missions (both *Euclid* and *WFIRST-2.4*) go further than just enabling higher-precision photometric redshifts due to a greater number of bands. The 2.4m aperture of the updated *WFIRST* satellite and its position above the atmosphere will provide estimated resolution of 0.11-0.14 arcseconds; comparison to ground-based imaging will enable studies of shape systematics, star-galaxy separation, and blending effects on photo-$z$'s (Spergel et al. 2013). For both satellite missions, the addition of bands in the NIR will improve photometric redshift performance, particularly at z>1.4 where the 4000 Å break moves beyond the $y/z$-band overlap, and dramatically reduce the rate of catastrophic redshift failures. The planned *WFIRST-2.4* high latitude survey plan covers 2000 deg$^2$ in four bands to depths comparable to full-depth LSST (5$\sigma$ point source depths of Y=26.7, J=26.9, H=26.7, F184=26.2), while *Euclid* plans to cover 15,000 deg$^2$ (spread over both hemispheres) in YJH to a depth of ∼24$^{\text{th}}$ magnitude (Laureijs et al. 2011). The addition of *Euclid*-like near-infrared data to simulated LSST photometry reduces the overall photo-$z$ uncertainty, catastrophic outlier rate, and bias all by a factor of ∼2 at z>1.5 (LSST Collaboration 2009). As overlapping area improves the utility of all datasets, coordinating the footprints of LSST, *Euclid*, and *WFIRST-2.4* should be done at the earliest stages. The limited common areal coverage can be used independently for studies that do not need full survey coverage, or the subset of galaxies with NIR photometry can be used to estimate and improve on the performance of overall LSST photo-z's, along the lines of the "super-photo-z's" described above. For some studies, the smaller areas and higher purity of photometric redshifts that combine ground- and space-based data may outweigh the smaller sky coverage (and hence sample size) that will be in common between multiple experiments.

*Quantifying the impact of template errors and incompleteness*   In template-based photo-$z$ algorithms, increasing the number of spectral templates increases both the computational requirements and the number of possible degeneracies, so the set is often limited to a small number (typically 6-30) of templates that approximately span the range of expected SEDs. Given that galaxies span a continuous distribution in properties, some will fall between the members of this finite set and will have slightly biased predicted redshifts, a contribution to the photo-$z$ errors sometimes referred to as "template mismatch variance". Determining the number and composition of an optimal template set is an open question that will affect data-processing requirements for Stage IV surveys.

In a similar manner, galaxies with SEDs not covered by the photo-$z$ template set will generally be fit using the "closest" matching SED, which most likely will have maximum likelihood at an incorrect redshift, introducing a bias to the photo-$z$ measurement of the unrepresented galaxy population. Abrahamse et al. (2011) investigated the impact of such template incompleteness with a toy model "leave one out" strategy to test the impact on redshift distributions for tomographic samples. Simulations of photo-$z$ measurements on mock LSST full-depth photometry, in which a set of ∼200 templates were used to generate the mock photometry but only a random half were used in photometric redshift determination, resulted in an average increase in photo-z scatter of 40% and an increase in bias of 50% compared to use of the full template set. These examples highlight the sensitivity of photo-$z$ results to the design of the template set, and underline why obtaining a comprehensive spectroscopic training sample is so valuable.

*Impact of blended objects:*   In DEEP2 data, the rate of objects that appear to be a single object from the ground but are resolved into multiple galaxies from space rises above 1% at z>1 (Newman et al. 2013). This is likely to set fundamental limits on photo-$z$ performance for ground-based surveys, as no single redshift and SED will describe the blended object. Since the LSST sample will be fainter and at higher redshifts, this will be a more severe problem for LSST than Stage III surveys.

*Complementarity between multiple photo-z estimators*   A variety of methods for estimating photometric redshifts exist. Different algorithms may predict different redshifts (or redshift distributions) for a given object, even though they rely on the same underlying photometric data. Given this situation, how to choose amongst (or combine the results of) multiple photometric redshift algorithms is an open problem.

One option is to combine redshift distributions from multiple algorithms in a manner that accounts for their covariance, but not choose between them directly; for instance, Dahlen et al. (2013) utilize hierarchical Bayesian techniques for this purpose. Likewise, Carrasco-Kind & Brunner (2013b) have developed a Bayesian approach that combines three different photometric redshift estimation methods, based on supervised learning, unsupervised learning, and template-fitting, which they find not only improves the overall estimation of the photometric redshift PDF but also aids in the identification of outliers. Further development in this area should prove beneficial in optimally combining different algorithms, each with their own systematics, to leverage the discord amongst photometric redshift estimates to provide additional useful information.

Figure 4-1, drawn from Gorecki et al. 2013, illustrates a different potential gain from combining multiple methods. Tests using CFHTLS data (Le Fevre et al. 2004, Newman et al. 2012, Garilli et al. 2008 and Lilly et al. 2007) found that photometric redshifts estimated using a template fitting method (denoted $z_P$ here) exhibited smaller scatter, while a training-set-based technique, denoted $z_{MLP}$ for Multi Layer Perceptron, exhibited fewer outliers with large differences between the true, spectroscopic redshift (dubbed $z_S$) and the photometric redshift. As seen in the figure, objects that exhibited a large value of $|z_P-z_{MLP}|$ in simulated data tend to have catastrophic redshift errors in the template-fit photo-z's (i.e., a large difference between fitted $z_P$ and true $z_S$). In this case, an improved photometric redshift could be obtained by using an MLP-based veto to exclude objects with questionable photometric redshifts, but otherwise use the higher-

precision template fitting technique.

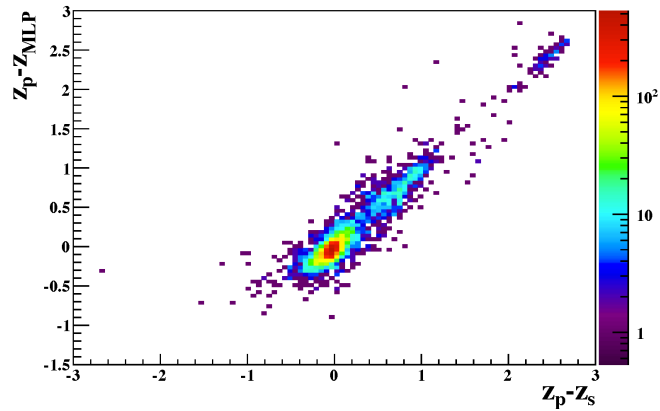

**Figure 4-1.** *Illustration of how the discrepancy between the template fitted photometric redshift estimate, $z_P$, and the neural network estimated photometric redshift, $z_{MLP}$, can be used to identify objects with problematic photo-z's (from Gorecki et al. 2013). In this simulated dataset, objects with large differences between the template-fit photo-z and the true, spectroscopic redshift ($z_s$) almost always also exhibit large differences between the photometric redshift estimated by two very different algorithms. Hence, one can use the concurrence between algorithms as a veto to reject objects with problematic photometric redshifts, yielding purer samples for dark energy studies.*

The disadvantage of the photo-z comparison approach is that it is difficult to describe formally. An interesting alternative, developed in Oyaizu et al (2008), is the nearest-neighbor error estimator, NNE[1]. As that work shows, NNE approaches can be very efficient at identifying outliers. Another interesting possibility for combining template and training set concepts is through developing simulations which forward-model the observed data, matching all observables (including not just imaging properties, but also the properties of the spectroscopic training set, including redshift completeness). For large data sets, these methods can provide enormous computational advantages relative to standard template-fitting methods, since tree-based techniques can be used to quickly search the observable space and derive redshift properties.

The variety of results showing that combining multiple algorithms can improve photo-z estimates suggests that optimal methods have yet to be developed. Ideally, one would like to be able to start with a theoretical model, incorporate the selection and noise properties of a given survey and obtain a set of simulated data that reproduces all observable properties (including photometry, clustering, and spectroscopic properties) for both imaging and training sets simultaneously. Reaching that ideal situation would require great improvement in our understanding of galaxy evolution; but such improvement (in our knowledge of distributions of color, luminosity, star formation history, spectral energy distributions, etc.) is already a major goal of research in that field. Large, highly complete spectroscopic datasets would enable great progress on this problem, while simultaneously improving algorithms for estimating individual photometric redshifts.

***Photometric Redshift PDF storage and Analysis*** Multiple studies have demonstrated that utilizing the full photometric redshift probability density function (PDF) rather than a single point estimate conveys significantly more information for cosmological analyses (see, e.g., Mandelbaum et al. 2008; van

---

[1] Available for download at (http://www.stanford.edu/~ccunha/nearest/ or in the github.com repository 'probwts'.

Breukelen & Clewley 2009). Several photometric redshift algorithms are now currently available that provide a full PDF (e.g., Benitez 2000; Ilbert 2006; Cunha et al. 2009; and Carrasco-Kind & Brunner 2013a). Using and storing this extra information for large imaging datasets remains an area where additional work remains to be done, especially if multiple PDF techniques will be combined into a meta-probability density function. Some work has been done recently (Carrasco-Kind, Chang, & Brunner 2013) to identify how PDFs can be represented efficiently to reduce storage space and improve computational speeds and data transfer rates, while minimizing the impact on cosmological measurements. Given the large data volume of Stage IV surveys such as LSST, however, additional work will be required to understand the effects of PDF resolution and dynamic range in different storage schemes on the accuracy of cosmological measurements.

## 5.1 Summary

In order for Stage III and Stage IV imaging-based dark energy experiments to achieve system-limited performance, large sets of objects with spectroscopically-determined redshifts will be required. The purposes of these samples are twofold:

- *Training:* Objects with known redshift are needed to map out the relationship between object color and $z$ (or, equivalently, to determine empirically-calibrated templates for the restframe spectral energy distributions of galaxies). This information is critical for all photometric probes of dark energy. The larger and more complete this "training set" is, the smaller the RMS photo-z errors will be, increasing the constraining power of imaging experiments.
- *Calibration:* For every subsample used in dark energy analyses (e.g., bins in photometric redshift), redshift distributions (or equivalently the distribution of differences between estimated and true redshifts) must be known to high accuracy for dark energy constraints not to be weakened. For instance, the uncertainty in the mean and RMS of the redshift distribution for subsamples must be ∼0.002(1+z) or less for Stage IV experiments to not be systematics-limited. Calibration may be done with the same spectroscopic data set used for training if it is extremely high in redshift completeness (i.e., no populations of galaxies to be used in dark energy analyses are systematically missed). Highly-accurate calibration is a necessity for all imaging experiments.

Essentially, training is the process of making the first several moments of the distribution of differences between estimated and true redshift as small as possible, improving photometric redshift precision; while the goal of calibration is to determine the actual values of those moments well, ensuring accuracy in our understanding of the photometric redshifts used.

It is most likely that these needs will be fulfilled by a variety of spectroscopy:

- *Training:* Spectroscopic redshift measurements for ∼30,000 objects over >∼15 widely-separated regions each at least ∼20 arcmin in diameter, down to the full survey depths, will likely be necessary for Stage IV surveys. Larger samples with higher completeness will allow improved photo-z templates/algorithms and reduce photo-z scatter, making them highly desirable. Table 2-2 presents the total time required for strawman training samples when utilizing a wide variety of instruments that are currently in development.
  - To attain high redshift success rates, which will be critical both for utilizing training sets for precision calibration and for enabling usage of the full redshift span of Stage IV surveys, near-infrared spectroscopy for large samples will also be required; this is best done from space.
- *Calibration:* If extremely high (>∼ 99.9%) completeness is not attained in training samples (existing deep redshift samples fail to yield secure redshifts for 30-60% of targets, so this is a likely possibility) the best known option for calibration of photometric redshifts is cross-correlation-based techniques. The proposed stage III spectroscopic BAO experiment eBOSS would provide a sufficient dataset for basic calibrations, particularly for Stage III experiments; the stage IV DESI experiment would provide excellent calibration with redundant cross-checks. An extension of DESI to the Southern hemisphere would provide the best possible calibration.

In addition to scoping out the details of and obtaining these training and calibration samples, there remains additional work to be done on photometric redshift methods to prepare for data from Stage III and Stage IV surveys. For training, key issues include investigation of to what degree high-precision, many-band multi-wavelength photometric redshifts may be used to train and calibrate algorithms for planned imaging experiments, including how to bootstrap information from deep infrared imaging (from *WFIRST-2.4* or *Euclid*) that cover limited area to wider-field surveys; characterizing the systematic effects of template mismatch and incompleteness; characterizing the impact of blended objects on ground-based samples; and the development of optimal methods for combining results from multiple photometric-redshift methods. For calibration via cross-correlation techniques, the key open issue is whether envisioned techniques for handling the evolution of the manner in which galaxies trace large-scale structure are sufficient to achieve Stage IV requirements; tests with existing data sets can address this question.

The authors gratefully acknowledge helpful discussions with G. Bernstein, S. Bridle, S. Dodelson, J. Green, B. Jain, S. Kahn, A. McConnachie, O. Le Fevre, and D. Spergel.